\documentclass[11pt,a4paper]{article}
\pdfoutput=1
\usepackage{jcappub}
\usepackage[utf8]{inputenc}

\usepackage{caption}
\usepackage[usenames,dvipsnames,svgnames,table]{xcolor}
\usepackage{gensymb}

\def\smica{{\tt SMICA}}
\def\nilc{{\tt NILC}}
\def\sevem{{\tt SEVEM}}
\def\commander{\texttt{Commander}}
\def\commonmask{\texttt{UPB77}}
\def\fdg{\hbox{$.\!\!^\circ$}}

\usepackage[T1]{fontenc}

\notoc

\begin{document}

\title{Constraints on direction-dependent cosmic birefringence from
\textit{Planck} polarization data}
\author[a,1]{Dagoberto Contreras,\note{Corresponding author.}}
\author[a,b]{Paula Boubel,}
\author[a]{and Douglas Scott}

\affiliation[a]{Dept. of Physics \& Astronomy, University of British Columbia, \\6224 Agricultural Road, Vancouver, BC, Canada}
\affiliation[b]{University of Guelph, \\50 Stone Rd E, Guelph, ON, Canada}

\emailAdd{dagocont@phas.ubc.ca}
\emailAdd{pboubel@mail.uoguelph.ca}
\emailAdd{dscott@phas.ubc.ca}

\abstract{Cosmic birefringence is the process that rotates the plane of
polarization by an amount, $\alpha$, as photons propagate through free space.
Such an effect arises in parity-violating extensions to the electromagnetic
sector, such as the Chern-Simons term common in axion models, quintessence
models, or Lorentz-violating extensions to the standard model. Most studies
consider the monopole of this rotation, but it is also possible for the effect
to have spatial anisotropies.  Paying particular attention to large scales, we
implement a novel pixel-based method to extract the spherical harmonics for $L
\le 30$ and a pseudo-$C_L$ method for $L > 30$. Our results are consistent with
no detection and we set 95\,\% upper limits on the amplitude of a
scale-invariant power spectrum of $L(L+1)C_L/2\pi < [2.2\, (\mathrm{stat.})\,
\pm 0.7\, (\mathrm{syst.})]\times10^{-5} = [0.07\, (\mathrm{stat.}) \pm 0.02\,
(\mathrm{syst.})] \,\mathrm{deg}^2$, on par with previous constraints. This
implies specific limits on the dipole and quadrupole amplitudes to be
$\sqrt{C_1/4\pi} \lesssim 0\fdg2$ and $\sqrt{C_2/4\pi} \lesssim 0\fdg1$, at
95\,\% CL, respectively, improving previous constraints by an order of
magnitude. We further constrain a model independent $M=0$ quadrupole in an
arbitrary direction to be $\alpha_{20} = 0\fdg02 \pm 0\fdg21$, with an
unconstrained direction. However, we find an excess of dipolar power with an
amplitude $\sqrt{3C_1/4\pi} = 0\fdg32 \pm 0\fdg10\, (\mathrm{stat.})\, \pm
0\fdg08\, (\mathrm{syst.})$, in the direction $(l, b) = (295^\circ, 17^\circ)
\pm (22^\circ, 17^\circ)\, (\mathrm{stat.})\, \pm (5^\circ, 16^\circ)\,
(\mathrm{syst.})$, larger than 1.4\,\% of simulations with no birefringence. We
attribute part of this signal to the contamination of residual foregrounds not
accounted for in our simulations, although this should be further investigated.}


\maketitle
\hfil
\flushbottom
\section{Introduction}

It is well known that our Universe violates parity via weak-sector
interactions. It is natural then to look for violations of parity in other
sectors. Here we use the cosmic microwave background (CMB) to constrain a
Chern-Simons type parity violation in the electromagnetic sector
\cite{Carroll1989vb}. In particular we focus on the effect of cosmic
birefringence, which is the in vacuo rotation of the plane of polarization of
photons. Such an effect would occur in modifications to electromagnetism or
from higher-dimensional operators in effective field theories, such as the
axion-photon coupling or in some quintessence models \cite{Carroll1998,
Lue1999}, or in models with new scalar degrees that are not quintessence
\cite{Li2008, Li2009, Pospelov2008, Finelli2008, Caldwell2011}, or models
driven by a modified gravitational interaction \cite{Feng2005}. In these
modifications, the addition of a term to the standard Lagrangian, which couples
a new pseudo-scalar field (or vector field) to the electromagnetic term
$F_{\mu\nu}\tilde{F}^{\mu\nu}$ (or $A_\mu\tilde{F}^{\mu\nu}$), would affect
left-handed photons and right-handed photons asymmetrically. This introduces a
phase shift difference between orthogonal polarization states that would
manifest itself as a rotation of the total linear polarization. A
spatially-varying pseudo-scalar field would result in variations of this
rotation angle, denoted by $\alpha$, across the sky. If such an effect were to
exist, then it would be lost in any search for isotropic $\alpha$. Hence it is
interesting to map out these potential fluctuations in $\alpha$.

A search for anisotropic birefringence is motivated because if, for example,
the scalar field is dynamical, then it should have spatial fluctuations that
would propagate as spatial variations in $\alpha$ \cite{Pospelov2008,
Li2008tma, Caldwell2011}. There are also models for which a uniform rotation
$\alpha$ vanishes, in which case these models would only be detectable in a
search for the anisotropies of $\alpha$. If detected, the power spectrum of
$\alpha$ would give considerable insight on the nature of the source of
birefringence and new physics. In particular the sourcing field could in
principle contain a special direction (similar to birefringent crystals such as
calcite or sapphire), which would impart a signal in a large scale map of $\alpha$. A final
reason for such a study is a more practical one, namely that a uniform $\alpha$
is degenerate with a systematic uncertainty in the orientation of the
detectors, particularly polarization-sensitive-bolometers (PSBs) used in CMB
experiments \cite{Leahy1997, Hu2002}. Currently, measurements of a uniform
rotation are systematics dominated at the $|\alpha| \lesssim 0\fdg3$ level
\cite{PlanckXLIX}. Searches for anisotropic $\alpha$ can therefore in principle
improve on this constraint, since a systematic uniform rotation would cancel
out, along with the monopole in $\alpha$.

The CMB is particularly suited for measuring cosmic birefringence because it is
polarized and because CMB photons propagate over cosmological distances essentially
unimpeded, where such a rotation could accumulate into a detectable signal. CMB
polarization is sourced by local density quadrupoles \cite{Hu1997_primer} at
the surface of last scattering, producing linear $Q$- and $U$-type
polarization. These quantities can be decomposed by their geometric properties
into $E$- and $B$-mode polarization components, which are the gradient (parity even) and
curl (parity odd) modes of the polarization field on the sky \cite{Seljak1997,
Kamionkowski1997, Hu1997_totangmom}. Under the assumption of parity
conservation the $T$--$B$ and $E$--$B$ correlations must be null, so that their
measurement informs us of parity violations.

Cosmic birefringence has previously been constrained using CMB anisotropies
from several experiments, most recently with \textit{Planck} data, under the
assumption of a uniform rotation. These results were found to be consistent
with our expectation of no cosmic birefringence (see Ref.~\cite{PlanckXLIX} and
references therein). In this paper, we use the \textit{Planck} 2015 (PR2) data
to consider anisotropies in $\alpha$. While the possibility of an anisotropic
$\alpha$ has previously been addressed using data from other experiments
\cite{Gluscevic2012, Gruppuso12, Li2013vga, Li2014oia, Lee2014rpa, Ade2015, B2K},
\textit{Planck} data can provide more stringent constraints, particularly on
the largest angular scales. Along with the $\alpha$ power spectrum, we
therefore constrain special directions in $\alpha$ taking the form of a dipole
or an $M=0$ quadrupole in an arbitrary coordinate system.

This paper is organized as follows.  In Section~\ref{sec:model} we explain the
effect cosmic birefringence has on the CMB angular power spectra. In
Section~\ref{sec:data} we describe the data and simulations used. In
Section~\ref{sec:methods} we describe our new map-space method used to estimate
the angle $\alpha$ locally as a function of direction on the sky. In
Section~\ref{sec:test} we demonstrate the effectiveness of our estimator on a
known input signal. In Section~\ref{sec:results} we present the results for our
baseline analysis pipeline. In Section~\ref{sec:systematics} we search for possible
sources of systematic effects that might affect our results, and we finally
conclude in Section~\ref{sec:conclusions}.

\section{Impact of birefringence on the CMB}
\label{sec:model}

A model of cosmic birefringence can be generated by including the following
term in the electromagnetic Lagrangian:
\begin{align}
  \mathcal{L} &= - \frac{\beta}{4M} \phi F_{\mu\nu} \tilde{F}^{\mu\nu} -
  V(\phi)
  \simeq \frac{\beta}{2M} \partial_\mu \phi\, A_\nu \tilde{F}^{\mu\nu} -
  V(\phi),
  \label{eq:Lagrangian}
\end{align}
where $F_{\mu\nu}$ is the electromagnetic field strength tensor and
$\tilde{F}^{\mu\nu}$ its dual, $\beta$ is a dimensionless coupling constant,
$M$ is a suppressing mass (usually taken to be the Planck scale), and the
potential $V(\phi)$ depends on the details of the model. The $\simeq$ symbol
here denotes equality up to a total derivative that has no effect on dynamics.
The interaction in Eq.~\eqref{eq:Lagrangian} is exactly the form of the
axion-photon coupling, while, for $V = \mathrm{constant.}$ the symmetry $\phi
\rightarrow \phi + \mathrm{constant}$ suppresses couplings to other Standard
Model particles \cite{Carroll1998}. The coupling of $\phi$ to
$F^{\mu\nu}\tilde{F}_{\mu\nu}$ treats left- and right-handed photons
asymmetrically, leading to a rotation in the plane of polarization as photons
propagate in vacuo. The amount of rotation is determined by the total change of
the field $\Delta \phi$ along the photon travel path and is given by
\begin{align}
  \alpha = \frac{\beta}{4M} \Delta \phi. \label{eq:alpha}
\end{align}

The existence of an angle $\alpha$ that is non-zero would be reflected in the
Stokes $Q$ and $U$ polarization parameters, which would be modified as
\begin{align}
  Q' \pm iU' &= e^{\pm 2i\alpha} (Q \pm iU).
  \label{eq:qualpha}
\end{align}
This effect induces $T$--$B$ and $E$--$B$ correlations that are otherwise
expected to be zero, along with smaller modifications to the parity-conserving
correlations. Specifically, the observed power (primed quantities) in these
cross-correlations would be
\begin{align}
	C_{\ell}'^{TT} &= C_{\ell}^{TT} \label{eq:ClTT}, \\
	C_{\ell}'^{EE} &= C_{\ell}^{EE}\cos^{2} (2\alpha)+C_{\ell}^{BB}\sin^{2} (2\alpha) \label{eq:ClEE}, \\
	C_{\ell}'^{BB} &= C_{\ell}^{EE}\sin^{2} (2\alpha)+C_{\ell}^{BB}\cos^{2} (2\alpha) \label{eq:ClBB}, \\
	C_{\ell}'^{TE} &= C_{\ell}^{TE}\cos(2\alpha) \label{eq:ClTE}, \\
	C_{\ell}'^{TB} &= C_{\ell}^{TE}\sin (2\alpha)\label{eq:ClTB},\\
	C_{\ell}'^{EB} &= \frac{1}{2}( C_{\ell}^{EE}-C_{\ell}^{BB})\sin (4\alpha)\label{eq:ClEB},
\end{align}
where the unprimed $C_{\ell}^{XY}$ are the spectra that would be measured in
the case of no cosmic birefringence. We will assume that $C^{BB}_{\ell}$ is
negligible, since \textit{Planck} has no direct detection of $B$ modes.
Employing the small-angle approximation\footnote{Our power spectrum results
will demonstrate that this is a good approximation. In the event that this
approximation breaks down, one should interpret the power spectrum to be for
the quantity $\sin{(4\alpha)}/4$ (as explained in Ref.~\cite{Gluscevic2012}).}
(in $\alpha$) we see that the only modifications to the CMB power spectra
appear as non-zero $C'^{TB}_\ell$ and $C'^{EB}_\ell$:
\begin{align}
  C'^{TB}_\ell(\hat{\pmb{n}}) &= 2 \alpha(\hat{\pmb{n}}) C^{TE}_{\ell},\\
  C'^{EB}_\ell(\hat{\pmb{n}}) &= 2 \alpha(\hat{\pmb{n}}) C^{EE}_{\ell},
  \label{eq:ebspectrum}
\end{align}
where we have now allowed $\alpha$ to depend on direction. As shown in
Ref.~\cite{PlanckXLIX}, constraints on birefringence from \textit{Planck} are
driven by the $E$--$B$ correlation, so we primarily focus on
Eq.~\eqref{eq:ebspectrum}. This relation suggests that we can use local
measurements of the $E$--$B$ correlation to determine $\alpha$ as a function of
direction on the sky.

The correlations of Eqs.~\eqref{eq:ClTT}--\eqref{eq:ClEB} can be searched for
either in harmonic space or pixel space. In the spatial domain, one can look at
the correlations between temperature extrema and polarization to reveal
$T$--$B$ cross-correlations.  Similarly, correlations between $E$-mode extrema
and polarization reveal $E$--$B$ correlations.  Both approaches were used to
constrain an isotropic $\alpha$ in Ref.~\cite{PlanckXLIX}. To analyse
polarization data in the neighbourhood of extrema, or ``peaks,'' the modified
Stokes parameters, $Q_{r}$ and $U_{r}$ are used. This involves a transformation
to radial and tangential components centred on each peak as the origin.
Specifically, the value of $Q_{r}$ at an angular distance $\theta$ from a peak
is the radial ($<$ 0) and tangential ($>$ 0) component of the polarization with
respect to the peak. The $U_{r}$ component is non-zero if the polarization is
rotated by 45$\degree$ with respect to these directions. The specific
transformations are
\begin{align}
 Q_{r}(\theta)&=-Q(\theta)\cos (2\phi) - U(\theta)\sin (2\phi)\label{eq:Qr}, \\
 U_{r}(\theta)&=\phantom{-}Q(\theta)\sin (2\phi) - U(\theta)\cos (2\phi)
 \label{eq:Ur}.
\end{align}

The transformed Stokes parameters are calculated in the neighbourhoods of each
peak. The patterns are expected to have azimuthal symmetry, so the data can
be compared to the following theoretical predictions (derived in
Refs.~\cite{Komatsu2011, PlanckXLIX}):
\begin{align}
    \langle U_{r}^{T} \rangle (\pmb{\theta}) &=- 2\alpha\int \frac{\ell
    d\ell}{2\pi} B_{\ell}^{2} p_{\ell}^{2}
    (\overline{b}^T_{\nu}+\overline{b}^T_{\zeta}\ell^{2}) C_{\ell}^{TE}
    J_{2}(\ell\theta)\label{eq:Urt}; \\
    \langle U_{r}^{E} \rangle (\pmb{\theta}) &=- 2\alpha\int \frac{\ell
    d\ell}{2\pi} B_{\ell}^{2} p_{\ell}^{2}
    (\overline{b}^E_{\nu}+\overline{b}^E_{\zeta}\ell^{2}) C_{\ell}^{EE}
    J_{2}(\ell\theta)\label{eq:Ure}.
\end{align}
Here $\pmb{\theta}$ is a radial vector, with $\theta$ its magnitude, $B_{\ell}$
is a $10'$ beam applied to the data, and  $p_{\ell}$ is the pixel window
function at
\texttt{HEALPix}\footnote{\url{http://www.healpix.sourceforge.net/}}
\cite{Gorski2005} $N_{\text{side}}=1024$ resolution. The quantity $J_{2}$ is
the second-order Bessel function of the first kind, and
$\overline{b}^{T,E}_{\nu, \zeta}$ are bias parameters that arise from the
selection of peaks from a Gaussian field \cite{Bond1987, Desjacques2008}, which
are discussed and calculated in Refs.~\cite{Komatsu2011, PlanckXLIX}.

\section{Data and simulations}
\label{sec:data}

Our baseline results use the full mission and half-mission \emph{Planck}
\cite{PlanckI} data splits for the $E$, $Q$, and $U$ polarization
maps,\footnote{Available at \url{http://www.cosmos.esa.int/web/planck/pla}}
specifically using the \smica\ component-separation procedure  \cite{PlanckX,
PlanckIX}, chosen for its relatively low noise level in polarization; however,
we use \commander, \nilc, and \sevem\ maps to check for consistency. The maps
are provided at a \texttt{HEALPix} $N_{\rm side} = 1024$ resolution, smoothed
with a $10^\prime$ beam. Along with these maps we use the common polarization
mask \commonmask\, in union with a mask that covers missing pixels specific to
the half-mission data split \cite{PlanckIX}.\footnote{Though the areas affected
by missing pixels are inadequate for measuring $\alpha$, they nevertheless have
very little effect on our results.} For our high-multipole likelihood
(described in Section~\ref{sec:highl}) we eventually degrade our maps to an
$N_{\rm side} = 256$ resolution. We then define a new conservative mask that is
simply the original mask degraded, with all pixels that contain a masked pixel in
the original mask being set to zero.

It is worth recalling that in 2015 the \emph{Planck} collaboration released
polarization data with some known systematic effects still present.
Specifically, there are large angular artefacts in the data that have yet to be
remedied \cite{PlanckI}, temperature-to-polarization leakage effects at smaller
scales \cite{PlanckVII, PlanckVIII, PlanckIX}, and a noise mismatch between the
data and simulations \cite{PlanckXII}. We account for the large-scale artefacts
by using high-pass-filtered versions of the data (and simulations), explicitly
a cosine filter that nulls scales $\ell \le 20$ and transitions to unity at
$\ell = 40$ \cite{PlanckIX}. We note that the best-fit
temperature-to-polarization leakage model was removed from the data
\cite{PlanckIX} and had negligible effect on the uniform $\alpha$ constraints
\cite{PlanckXLIX}, though we do not explicitly test for its impact here. The
noise mismatch does not greatly affect our results, since our data come from
the $E$--$B$ cross-correlation; nevertheless we \emph{do} use auto-correlations
of $\alpha$ (which are dominated by the $E$--$E$ and $B$--$B$ correlations
\cite{Gluscevic2012}) on large scales in order to have a well defined
likelihood (see Section~\ref{sec:lowl}) and have checked that results obtained
are consistent with the cross-correlation of $\alpha$ determined by the
half-mission data.

We use a suite of simulations for which the power in $\alpha$ is null, in order
to estimate uncertainties for our power spectrum results. We generate
polarization simulations using the following fiducial cosmology, which is
consistent with the data \cite{PlanckXIII}:
$\omega_{\rm b} = 0.0222$;
$\omega_{\rm c} = 0.1203$;
$\omega_\nu = 0.00064$;
$\Omega_\Lambda = 0.6823$;
$h = 0.6712$;
$n_{\rm s} = 0.96$;
$A_{\rm s} = 2.09\times 10^{-9}$;
and $\tau = 0.065$.
Here $\omega_{\rm x} \equiv \Omega_{\rm x} h^2$ are the physical densities. We
add noise power to our simulations in order to match the total power in our
data maps; this, however, does not include a correlated noise component or
non-Gaussian foreground residuals. For our birefringence analysis the
cosmological parameters are fixed to the values reported above. This seems to
be a safe assumption, since $\alpha$ has no effect on $C^{TT}_{\ell}$ and only
affects $C_{\ell}^{TE}$, and $C_{\ell}^{EE}$ (and thus parameters) at second or
higher order (see Eqs.~\ref{eq:ClTT}--\ref{eq:ClEB}).
For small $\alpha$, the effect of lensing is orthogonal to an anisotropic
birefringence \cite{Gluscevic2012,Yadav:2009eb}, however a bias appears at the
power spectrum level \cite{Namikawa:2016fcz}. This bias on the power spectrum
is, however, sub-percent at \emph{Planck} noise levels \cite{Namikawa:2016fcz}
and therefore we ignore its effects here.
We further generate a
suite of simulations with a particular scale-invariant $\alpha$ power spectrum,
described in Section~\ref{sec:test}, to demonstrate the effectiveness of our
$\alpha$ reconstruction
and to determine the corresponding reconstruction bias.

\section{\boldmath Measuring $\alpha$ locally}
\label{sec:methods}

We use Eq.~\eqref{eq:Ure} to define an unbiased estimator for $\alpha$ at the
location of every peak $p$ ($\hat{\alpha}_p$):
\begin{align}
  \tilde{U}_{r} (\pmb{\theta}_p) &=- 2 \int \frac{\ell
  d\ell}{2\pi} B_{\ell}^{2} p_{\ell}^{2}
  (\overline{b}_{\nu}+\overline{b}_{\zeta}\ell^{2})
  C_{\ell}^{EE}J_{2}(\ell\theta); \\
  \hat{\alpha}_p &= \frac{\sum_p \hat{U}_r(\pmb{\theta}_p)
  \tilde{U}_r(\pmb{\theta}_p)}{\sum_p \tilde{U}_r(\pmb{\theta}_p) \tilde{U}_r(\pmb{\theta}_p)}.
  \label{eq:dirdependent}
\end{align}
Here $\hat{U}_r(\pmb{\theta}_p)$ is the value of the data, and $\pmb{\theta}_p$
is a radial vector centred at the location of peak $p$. The peak positions are
determined by the full-mission $E$-mode map, while the above fit is performed
on the full or half-mission $Q$ and $U$ maps. Equation~\eqref{eq:dirdependent}
is a simple linear least-squares fit to $\tilde{U}_r$ with the identity as the
covariance matrix for $\alpha$ and the sum is performed over all
\emph{unmasked} pixels within a $2^\circ$ radius (chosen since the
$\tilde{U}_r$ profile vanishes at distances $> 1\fdg5$, as demonstrated in
Refs.~\cite{Komatsu2011, PlanckXVI, PlanckXLIX}). We also remove the monopole
in $\alpha$, to suppress any leakage to higher multipoles, since the monopole
is systematics dominated and large \cite{PlanckXLIX}; however, we check that
this step has no significant effect on our results.

We fit for a scale-invariant power spectrum, which takes the form
\begin{align}
  \frac{L(L+1)}{2\pi} C_L &\equiv A,
  \label{eq:scaleinvariant}
\end{align}
for a constant $A$. By convention we refer to $\alpha$ multipoles as ``$L$'' to
distinguish them from $\ell$ multipoles (for the temperature and polarization
anisotropies). This spectrum would be realized in a model
containing nearly massless psuedo-scalar degrees of freedom coupled to photons,
such as in Ref.~\cite{Caldwell2011}, for $L \lesssim 100$. A model like this is
constrained extremely well at low $L$ compared to high $L$, which puts
\emph{Planck} data at a distinct advantage compared to smaller coverage
(although
with higher sensitivity) ground-based experiments. For this reason we pay particular
attention to recovering the low $L$s accurately (see Section~\ref{sec:lowl}).

Previous estimators of anisotropic $\alpha$ in the literature compute the
contribution to the 4-point function (of $TB$ or $EB$) in harmonic space using
standard quadratic maximum likelihood techniques \cite{Gluscevic2012} (similar
to CMB lensing techniques), or using the 2-point correlation function
\cite{Li2013vga, Li2014oia}. Our approach reconstructs the 4-point function by
simply measuring the variation of the 2-point function locally in the data. In
the following subsections we describe how we take the local measurements of
$\alpha$ and compute maps at low and high resolution. For power spectra it is
worth recalling that the auto-spectra of an $\alpha$-map determined by the
$E$--$B$ correlation is \emph{not} the power spectrum of $\alpha$. This is
because the auto-spectrum necessarily contains contributions from the $E$--$E$
and $B$--$B$ correlations that are non-zero even if $\alpha = 0$
\cite{Gluscevic2012}. One can obtain the true power spectrum by subtracting the
mean power spectrum from simulations with a null $\alpha$ spectrum or by using
cross-correlations; we employ both of these methods below.

\subsection{Low multipoles}
\label{sec:lowl}

We begin by taking the $\alpha_p$ values found above and using a pixel fit to
recover the spherical harmonics $\alpha_{LM}$,
\begin{align}
  \tilde{\alpha}_{LM} &= \frac{\sum_p w_p \tilde{\alpha}_p Y^*_{LM}(\theta_p,
  \phi_p)}{\sum_p w_p |Y_{LM}(\theta_p, \phi_p)|^2}.
  \label{eq:lowlmethod}
\end{align}
Here the weights $w_p$ are uniform per pixel or chosen to incorporate the
uneven hits distribution from the \emph{Planck} scanning strategy. For our
baseline we use a uniform weighting scheme $w_p=1$; however, we also consider a
weighting scheme where $w_p$ is given by a smoothed version of the 217-GHz hits
map\footnote{The map is smoothed with a $2^\circ$ top-hat beam, chosen to match
our method of Section~\ref{sec:methods} that fits for $\alpha$ over pixels
within $2^\circ$ of each peak.} (denoted $H_p$, which we plot later). Note that
our simulations do not contain the effects of the \emph{Planck} scan strategy
and therefore only uniform weighting is used for them. This method accounts for
the mask by simply not using any masked pixels, i.e., all $\alpha_p$ values
come from \emph{unmasked} areas. We then compute the power spectrum of $\alpha$
either by taking the cross-correlation of $\alpha^1_{LM}$ and $\alpha^2_{LM}$,
determined with the half-mission 1 and 2 maps, or by taking the auto-spectrum
of $\alpha_{LM}$ determined by the full-mission data and subtracting the mean
of the $\alpha_{LM}$ auto-spectra calculated from null simulations. The latter
method is in principle more sensitive to the noise properties of the data, but
we find it to be consistent with the former method, and it has the advantage of
allowing us to form a Gaussian likelihood for the $\alpha_{LM}$s when fitting
for a model.
For this reason we only display these auto-correlation spectra in our low $L$
results that follow.

A direct calculation of Eq.~\eqref{eq:lowlmethod} is computationally expensive
for large $L_{\max}$, so we limit this approach to $L_{\max} = 30$ and consider
higher multipoles only in the following subsection. The low-$L$ likelihood then takes
the form
\begin{align}
  -2\log \mathcal{L} (\{\tilde{\alpha}_{LM}\}_{L=1}^{30}|A) &= \log |{\sf V}| +
  \sum^{L=30}_{LML'M'} \tilde{\alpha}_{LM}^{\vphantom{-1}}\,
  {\sf V}^{-1}_{LML'M'}\,
  \tilde{\alpha}^{*\vphantom{-1}}_{L'M'} + \mathrm{constant}, \\
  {\sf V}_{LML'M'} &= \left< C_{L} \right> \delta_{LL'} \delta_{MM'} +
  \mathcal{B} A
  \frac{2\pi}{L(L+1)} \delta_{LL'} \delta_{MM'}.
  \label{eq:lowllike}
\end{align}
The average is taken over a suite of simulations with $A=0$.
$\mathcal{B}$ is a normalizing factor to ensure that our estimator is unbiased,
and is determined
by cross-correlating the input $\alpha$ realizations with their corresponding
measured value from a reference power spectrum.
Bayes' theorem,
along with a flat prior on $A$ (specifying $A \ge 0$), allows us to turn this
into a likelihood for $A$ given the data and hence to obtain the posterior for
the amplitude of a scale-invariant spectrum.

\subsection{High multipoles}
\label{sec:highl}

\begin{figure}[ht!]
\begin{center}
\mbox{\includegraphics[width=0.6\hsize]{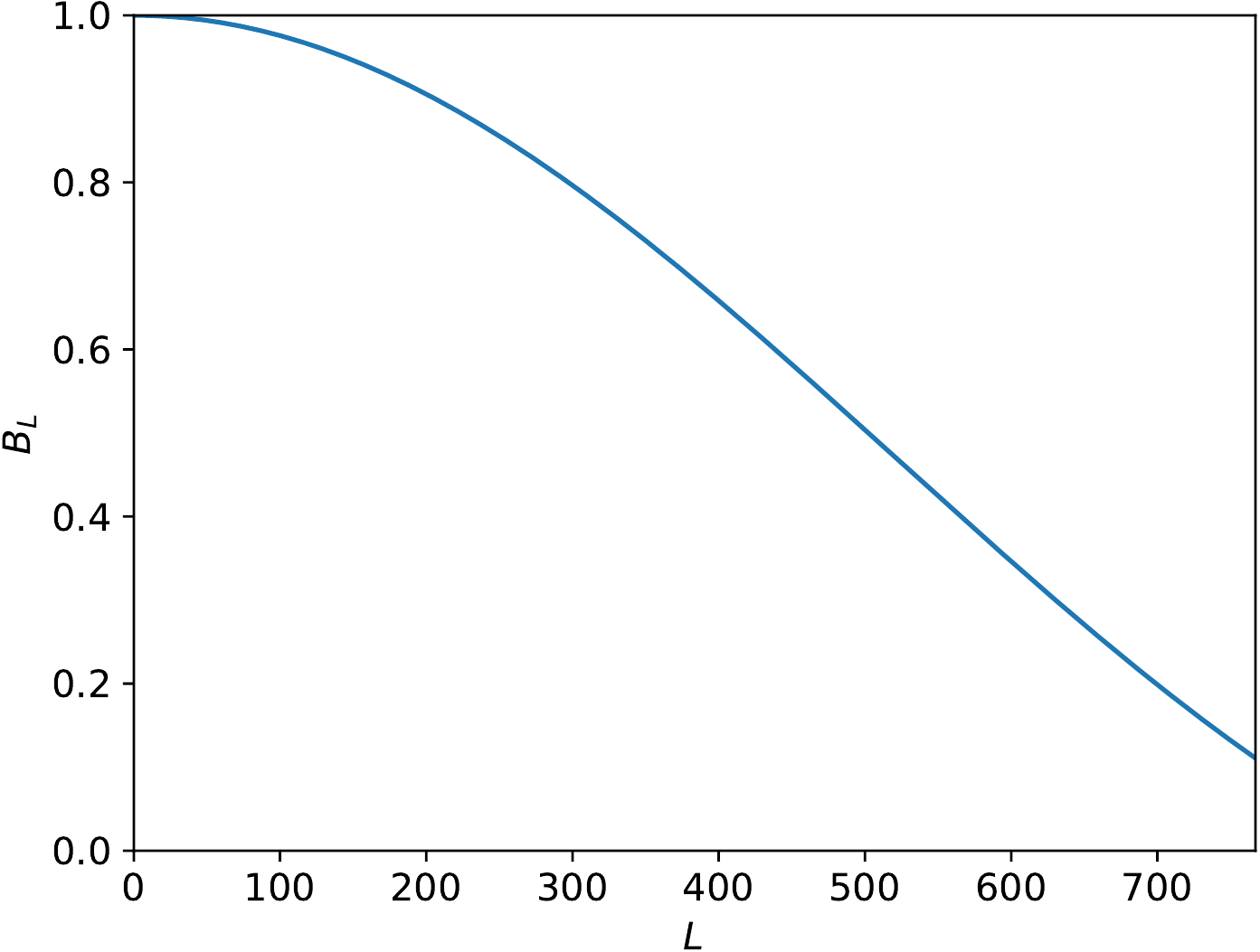}}
\end{center}
\caption{Effective beam for the high-$L$ analysis described in
Section~\ref{sec:highl}, induced by the smoothing procedure.}
\label{fig:beam}
\end{figure}

To produce our high-$L$ map we need to apply a smoothing to our
$\tilde{\alpha}_p$ values. The mean separation between peaks is about $0\fdg2$,
and therefore we define an $N_{\rm side} = 256$ map (denoted
$\tilde{\alpha}^{256}_p$) populated with the mean of $\tilde{\alpha}_p$ over
all pixels within a radius of $0\fdg25$:
\begin{align}
  \tilde{\alpha}^{256}_p &= \frac{\sum_{p' \in |p-p'|\le 0.25^\circ}
  \tilde{\alpha}_{p'}}{\sum_{p'
  \in |p-p'| \le 0.25^\circ}}.
  \label{eq:highlmap}
\end{align}
This procedure induces a beam function ($B_L$), which we show in
Fig.~\ref{fig:beam}. Our high-$L$ power spectrum is the cross-correlation of
Eq.~\eqref{eq:highlmap} between half-mission~1 and half-mission~2 data, with a
correction for the beam and the cut sky, i.e.,
\begin{align}
  \tilde{a}_{LM} &= \int d\Omega \tilde{\alpha}^{256}_p M(\Omega) Y^*_{LM}, \\
  \tilde{C}^{\alpha_1\alpha_2}_L &= \left< \tilde{a}^1_{LM} \tilde{a}^{*2}_{LM}
  \right> \approx f_{\rm sky} B^2_{L} p^2_{L} C^{\alpha_1\alpha_2}_L,
  \label{eq:psuedocl}
\end{align}
where $f_{\rm sky}$ is the fraction of the unmasked sky, $M(\Omega)$ is the
applied mask, $B_L$ is the effective beam function (see Fig.~\ref{fig:beam})
induced by the smoothing procedure, and $p_L$ is the pixel window function
specific to an $N_{\rm side} = 256$ resolution map. The approximation in
Eq.~\eqref{eq:psuedocl} is the exact \texttt{MASTER} \cite{Hivon2001}
correction due to the masking for an $L$-independent power spectrum for which
the data and simulations are consistent. We further bin the power spectrum in
bins of size $\Delta L = 50$ (with $L_{\min} = 31$ so as not to double count
the low-$L$ data), to minimize correlations induced by the mask from
neighbouring $L$ modes.

To fit for a scale-invariant power spectrum (Eq.~\ref{eq:scaleinvariant}) we
form $\chi^2$ as a function of $A$:
\begin{align}
  \chi^2 &= \sum_{bb'} \left(C^{\alpha_1\alpha_2}_b - C^A_b\right)
  {\sf G}^{-1}_{bb'}
  \left(C^{\alpha_1\alpha_2}_{b'} - C^A_{b'}\right),
  \label{eq:highlchi2}
\end{align}
where ${\sf G}^{-1}_{bb'}$ is the binned inverse covariance matrix derived from
simulations (generated with $A=0$) and $C^A_b$ is the model power spectrum
(Eq.~\ref{eq:scaleinvariant}) binned in the same way as the data and
simulations. We have verified with our simulations that each bin is close to
Gaussian and so we form a likelihood as
\begin{align}
  \mathcal{L} (\{C^{\alpha_1\alpha_2}\}_b | A) &\propto \exp{(-\chi^2/2)}.
  \label{eq:highllike}
\end{align}
Once again we can transform this into a likelihood for $A$ given the data, with
a flat prior on $A$.

Finally we combine our low-$L$ and high-$L$ likelihoods to form a joint
constraint on $A$ by simply taking the product of the likelihoods in
Eqs.~\eqref{eq:lowllike} and \eqref{eq:highllike}. This assumes that the data
at low $L$ are uncorrelated with the data at high $L$. This is known to not be
strictly true, since the mask will induce correlations, particularly for $L=30$
with the first bin of the high-$L$ data (which uses $L_{\min} = 31$). However,
this small correlation has very little impact on our results and the difference
in the results
is minimal if $L_{\min}$ is increased to reduce the correlation.

\section{Tests of the method}
\label{sec:test}

\begin{figure}[ht!]
\begin{center}
\mbox{\includegraphics[width=0.495\hsize]{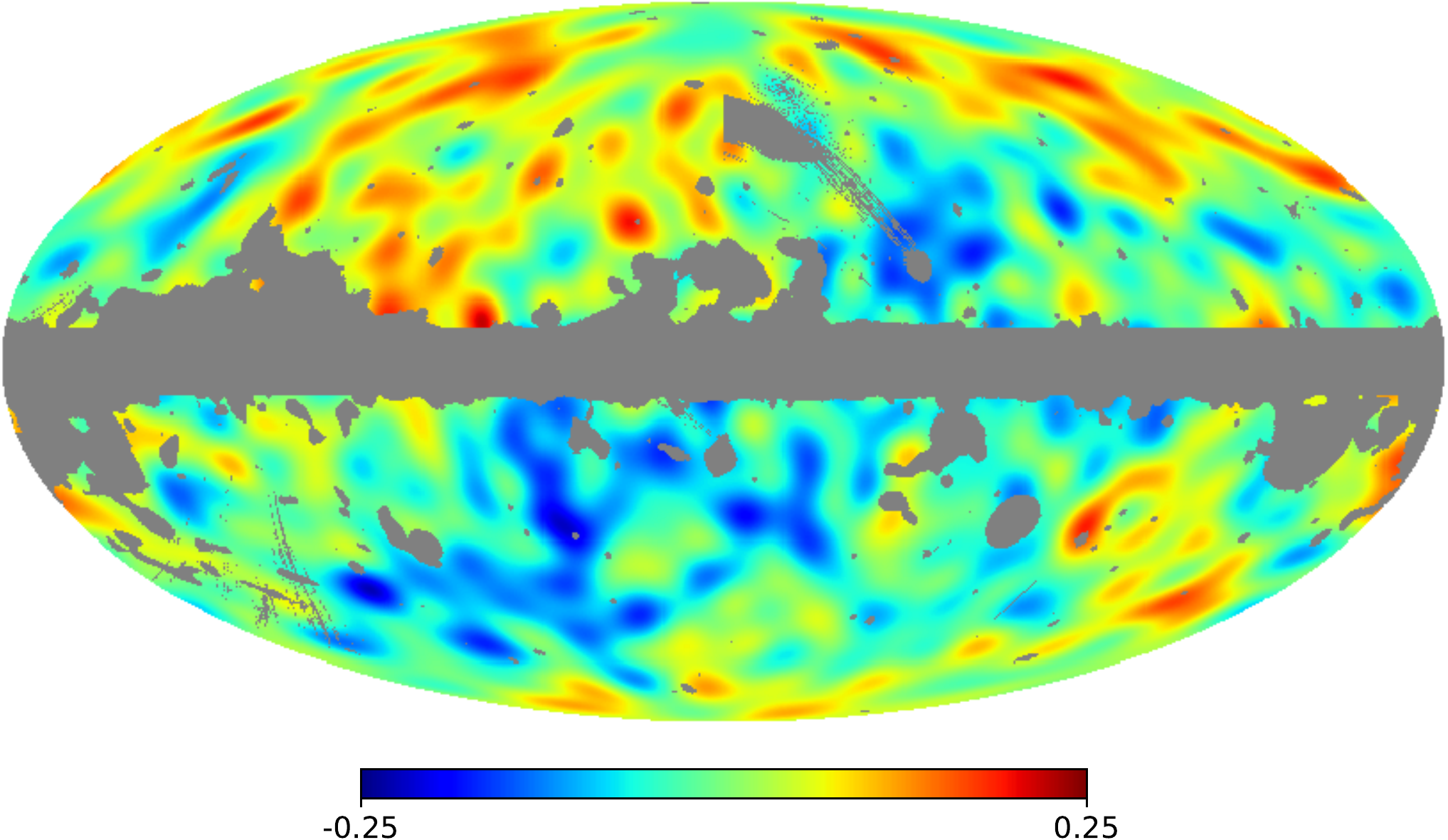}}
\mbox{\includegraphics[width=0.495\hsize]{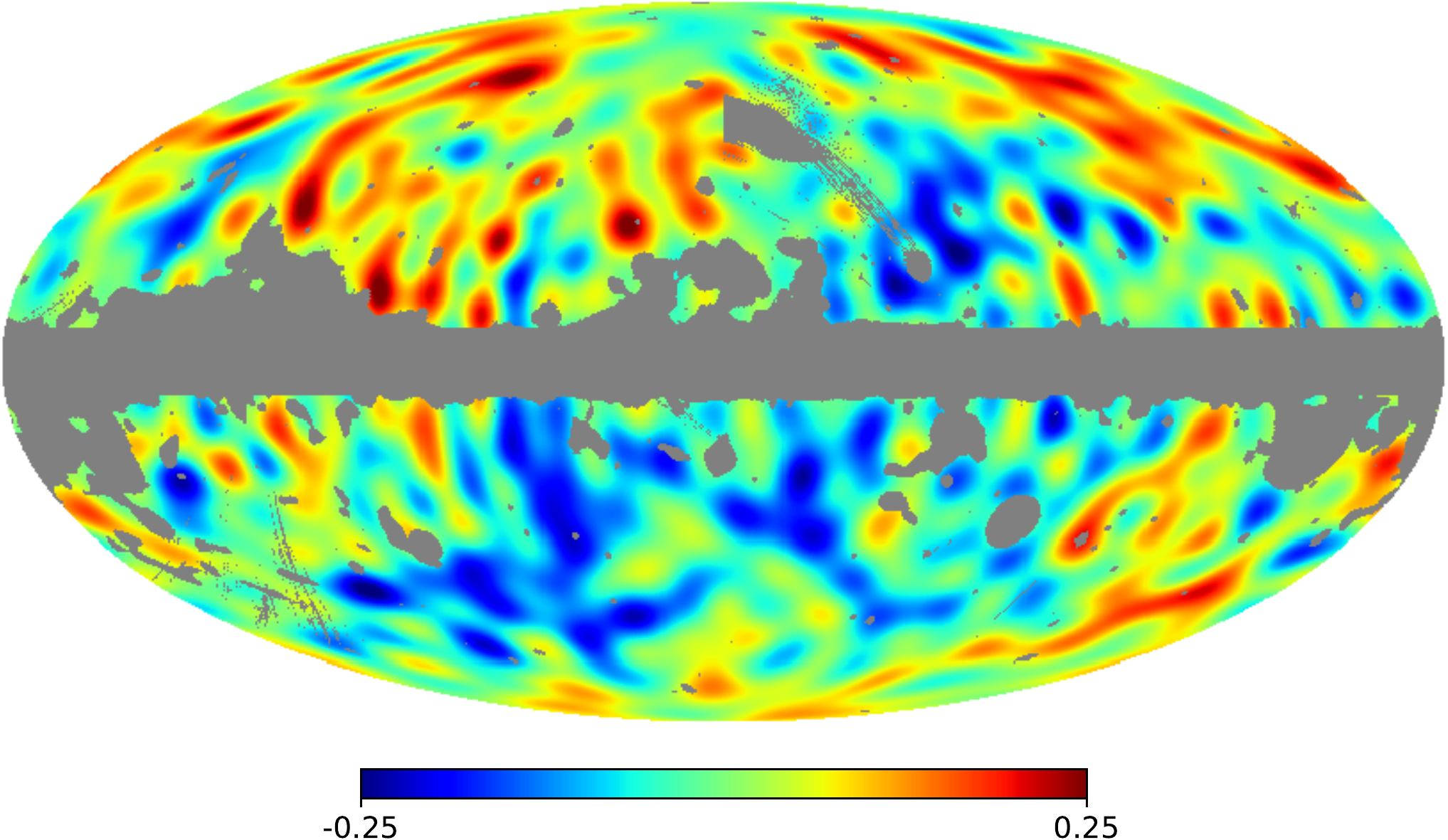}}
\mbox{\includegraphics[width=0.495\hsize]{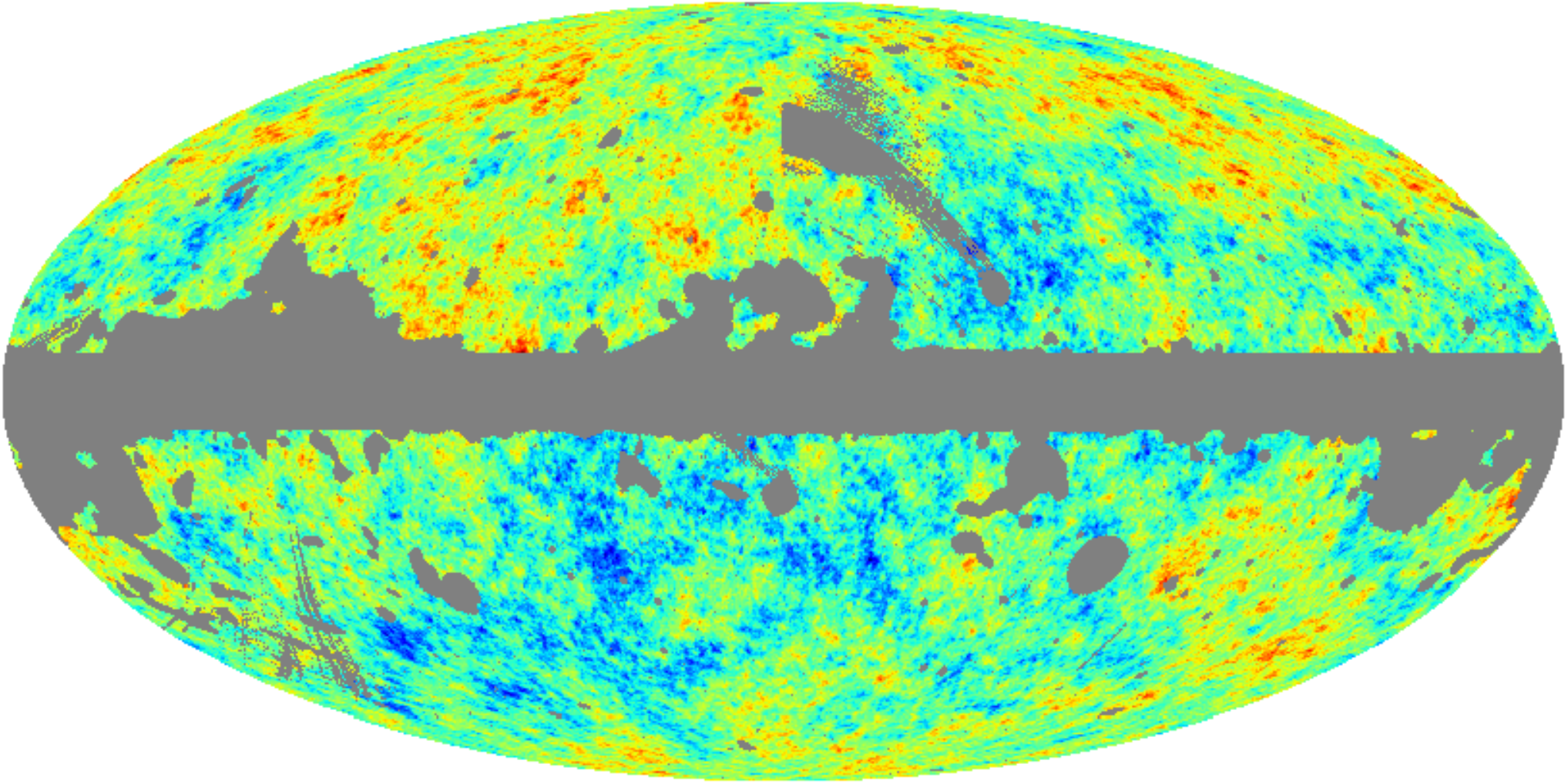}}
\mbox{\includegraphics[width=0.495\hsize]{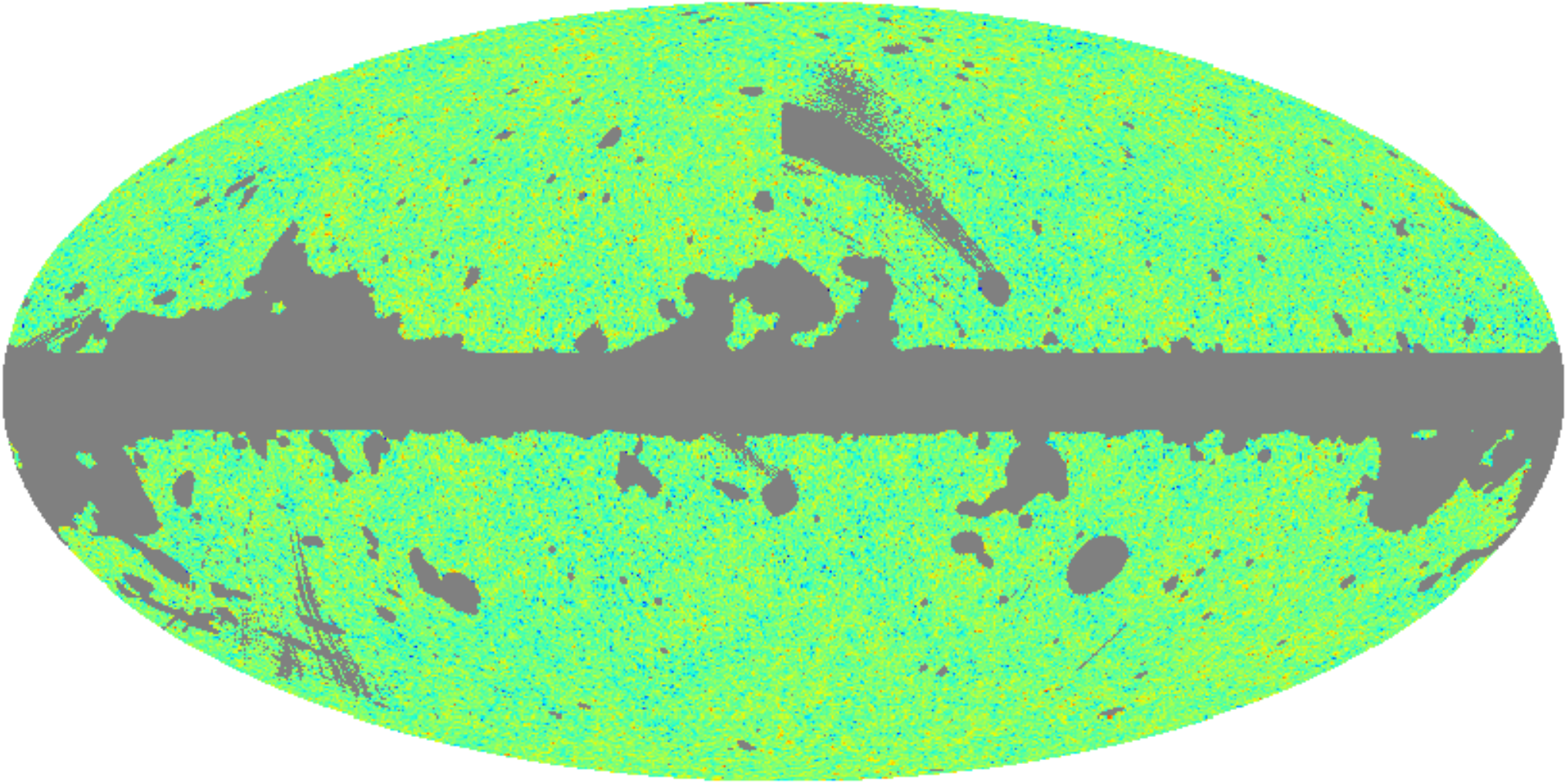}}
\end{center}
\caption{{\it Top}: Low-$L$ ($1 \leq L \leq 30$) $\alpha$-maps for an input
$\alpha$ realization (left) and for reconstruction by our method, as described
in Section~\ref{sec:lowl} (right). {\it Bottom}: High-resolution $\alpha$-maps
for an input $\alpha$ realization (left), along with the Weiner-filtered output
of our high-$L$ reconstruction, as described in Section~\ref{sec:highl} (right).
The induced beam (Fig.~\ref{fig:beam}) is applied to the input map for
comparison purposes. The input and output maps are clearly correlated, although
the output has considerably more scatter on small scales due to the significant
noise in the polarization maps.}
\label{fig:inputoutput}
\end{figure}

We first test the method described above in Section~\ref{sec:methods} by
generating a single simulation with a known realization of $\alpha$ from an
input power spectrum. To do this we generate a $Q$ and $U$ realization from the
cosmology described in Section~\ref{sec:data}, and we then modify the $Q$ and
$U$ maps by the relation Eq.~\eqref{eq:qualpha}. In this test the $\alpha$ map
is a realization of a scale-invariant power spectrum with $A = 10^{-2}/2\pi$
(chosen for visualization purposes), and is shown in Fig.~\ref{fig:inputoutput}
(left panels). We further input a noise realization to our simulation, so that
the total power in each of the $Q$ and $U$ maps is consistent with the data.

\begin{figure}[ht!]
\begin{center}
\mbox{\includegraphics[width=0.7\hsize]{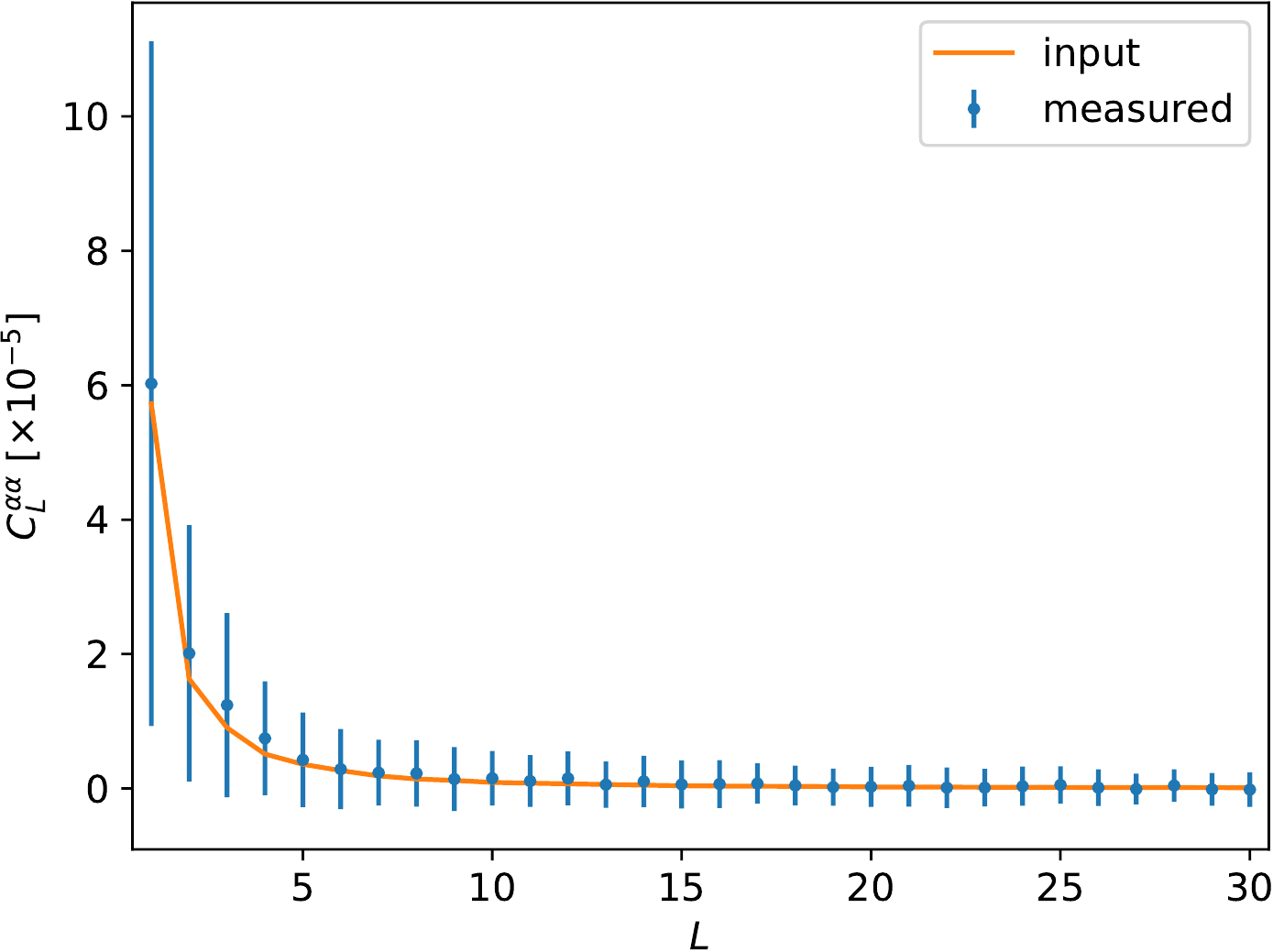}}
\end{center}
\caption{Recovery of a scale-invariant $\alpha$ power spectrum with $A =
10^{-4}/2\pi$ from a suite of simulations. The blue points are the mean
recovered power spectrum from simulations, while the bars denote the standard
deviations from the same set of simulations. The input power spectrum is shown
in orange. The overestimation comes from the reconstruction bias of the method,
which is accounted for by the $\mathcal{B}$ parameter in
Eq.~\eqref{eq:lowllike} in all subsequent power spectrum plots.}
\label{fig:inputcls}
\end{figure}

Upon applying our method of recovering $\alpha$, we obtain the panels on the
right-hand side of Fig.~\ref{fig:inputoutput}, for both low and high
resolution. At the level of the $\alpha$ maps the output of our analysis is
quite consistent with the input; however, there is considerably more noise in
our output maps (particularly at high resolution) due to the addition of
significant noise power. At the power spectrum level we also obtain very good
agreement with our input spectrum, with the scatter attributable to the noise
in the simulation. In Fig.~\ref{fig:inputcls} we show the mean recovered power
spectrum on a suite of simulations with $A = 10^{-4}/2\pi$, along with the
theoretical curve and the corresponding uncertainties. We find that our
reconstruction slightly overestimates the true power spectrum at the 30\,\%
level, which we correct for in the $\mathcal{B}$ parameter in
Eq.~\eqref{eq:lowllike}.

\section{Results}
\label{sec:results}

First we confirm that we recover the results for the $\alpha$-monopole from
Ref.~\cite{PlanckXLIX}.  However, as already noted, in the main analysis of
this paper we remove the monopole (which is dominated by systematic effects) so
that it does not leak into higher multipoles.

\subsection{Maps and power spectrum}
\label{sec:maps}

\begin{figure}[ht!]
\begin{center}
\mbox{\includegraphics[width=0.49\hsize]{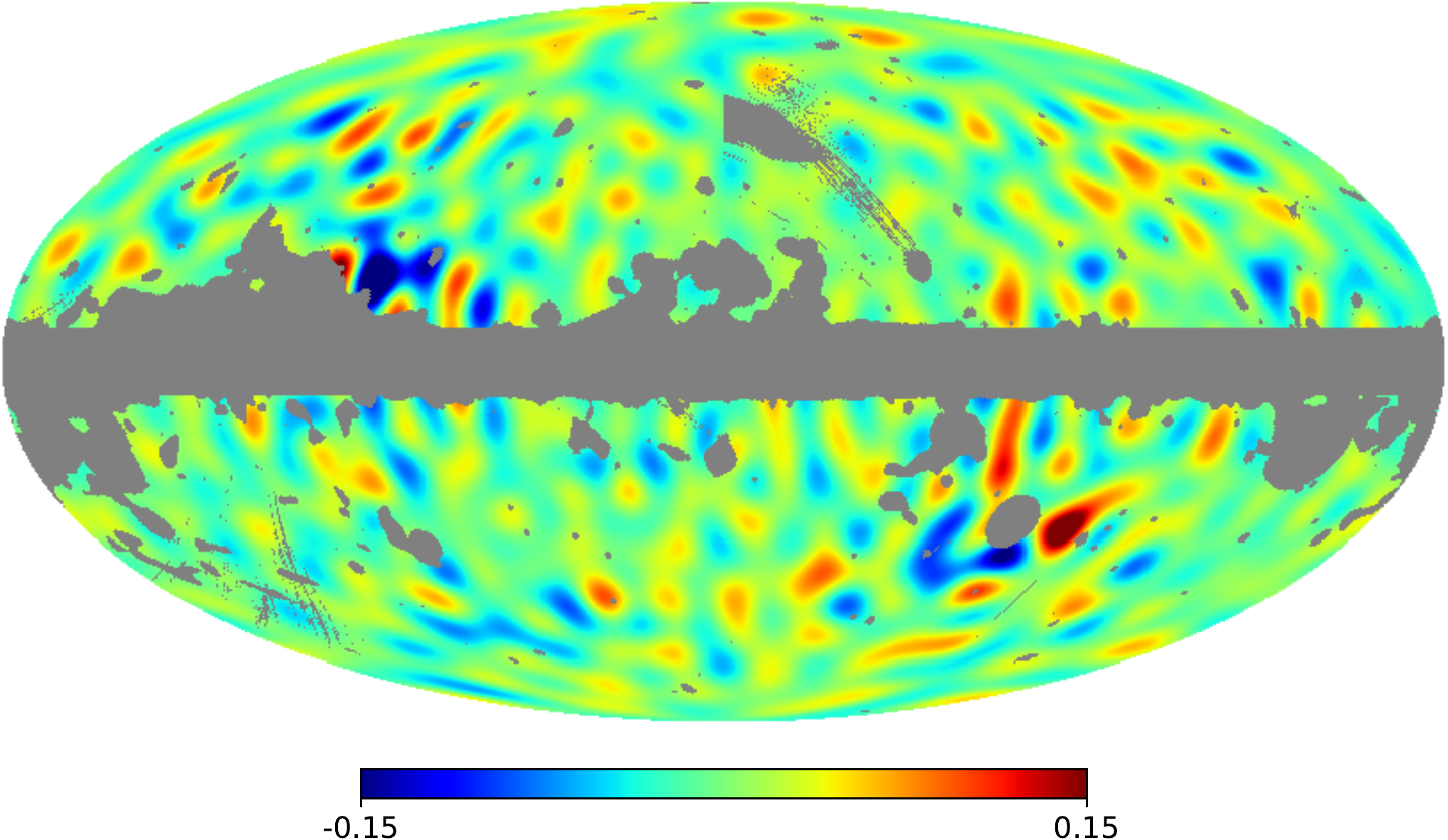}}
\mbox{\includegraphics[width=0.49\hsize]{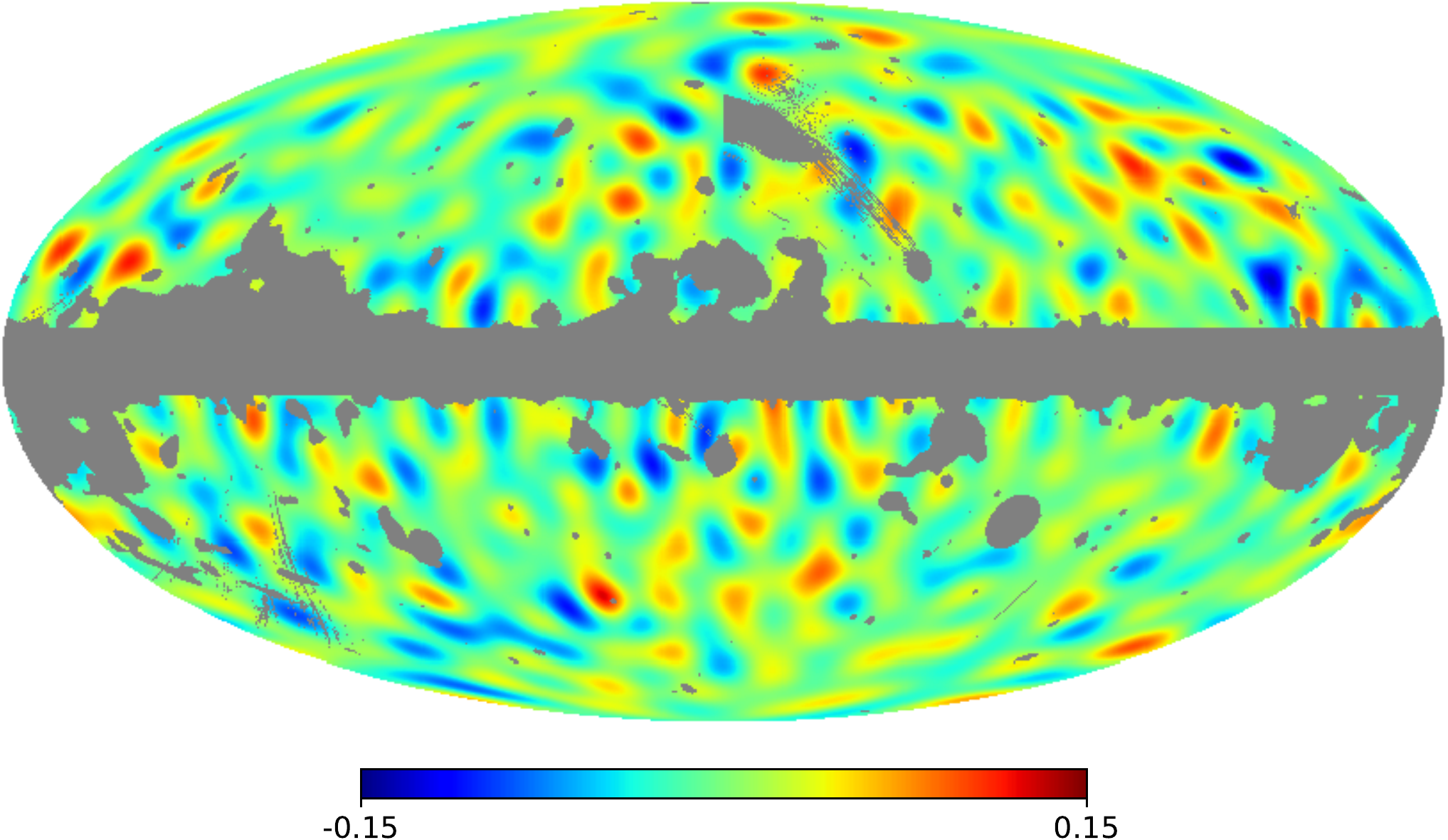}}
\mbox{\includegraphics[width=0.49\hsize]{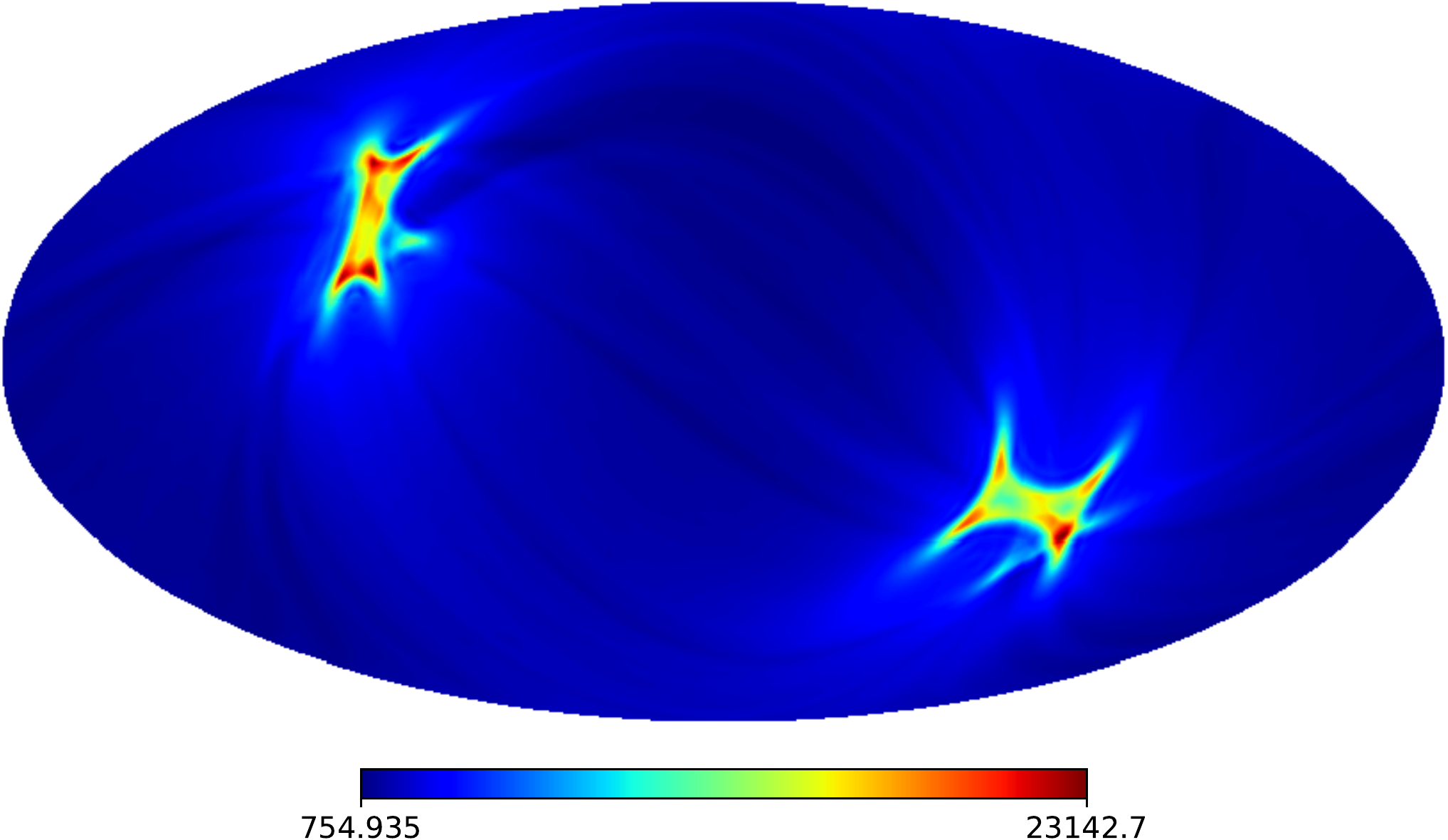}}
\mbox{\includegraphics[width=0.49\hsize]{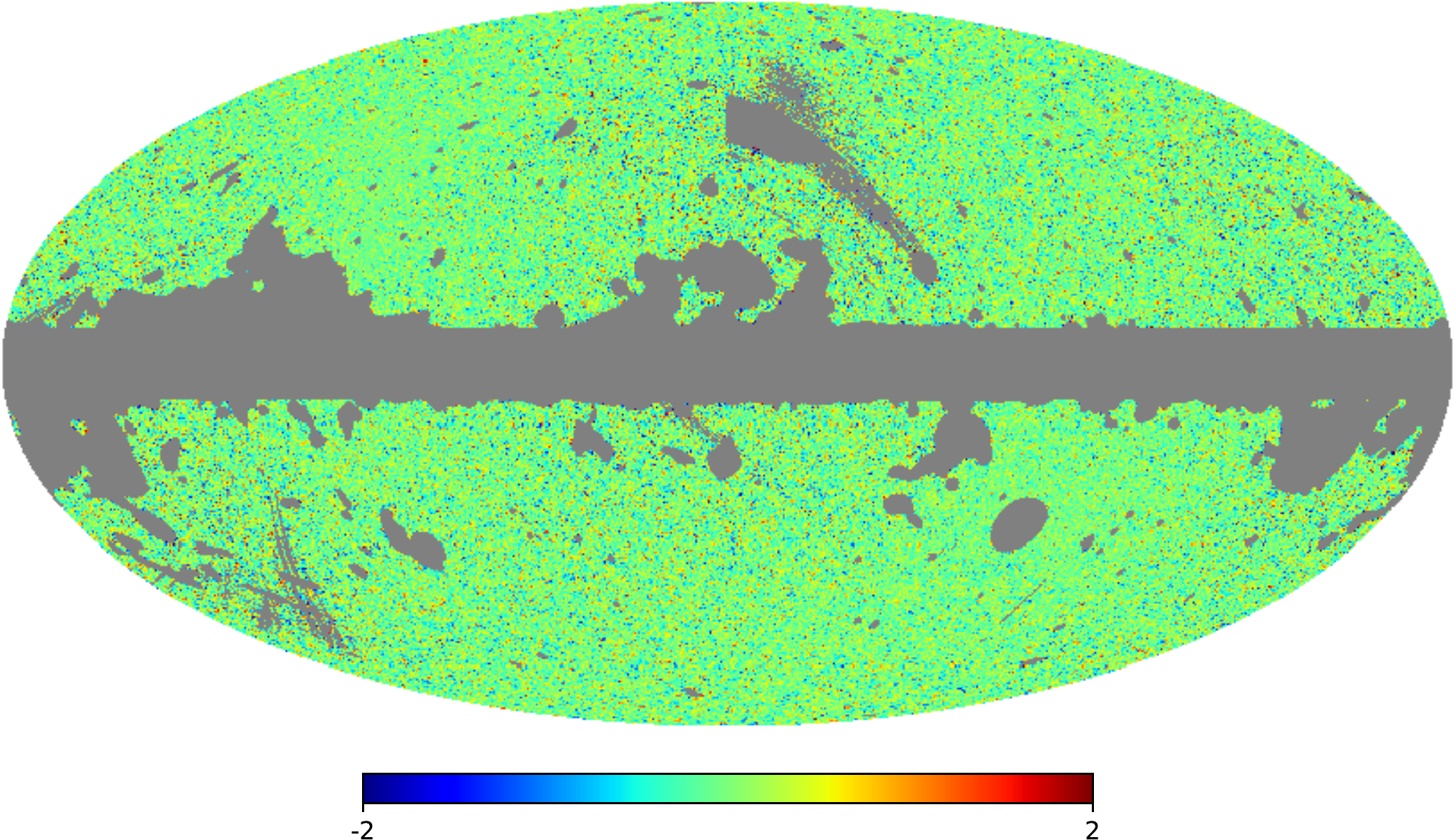}}
\end{center}
\caption{{\it Top}: Low resolution ($L_{\max} = 30$) data maps of $\alpha$
weighted by the hits map (left) or using uniform weighting (right). {\it
Bottom}: Smoothed hits map used for the $w_p = H_p$ analysis (left), together
with high-resolution data map (right).}
\label{fig:lowlmaps}
\end{figure}

Our low-$L$ and high-$L$ maps for the data are shown in the top row and bottom
right panels of Fig.~\ref{fig:lowlmaps}, respectively. Our two low-$L$ maps are
clearly strongly correlated; however, using the weighting given by the hits map
in Fig.~\ref{fig:lowlmaps} (bottom left) we see large-scale features near the
Ecliptic poles and Galactic plane. While, visually striking, these features
appear to have very little effect on our power spectrum results (to be
discussed in Section~\ref{sec:foregrounds}). They nevertheless point to
systematic effects associated with residual foregrounds not accounted for in
our simulations.

\begin{figure}[ht!]
\begin{center}
\mbox{\includegraphics[width=0.53\hsize]{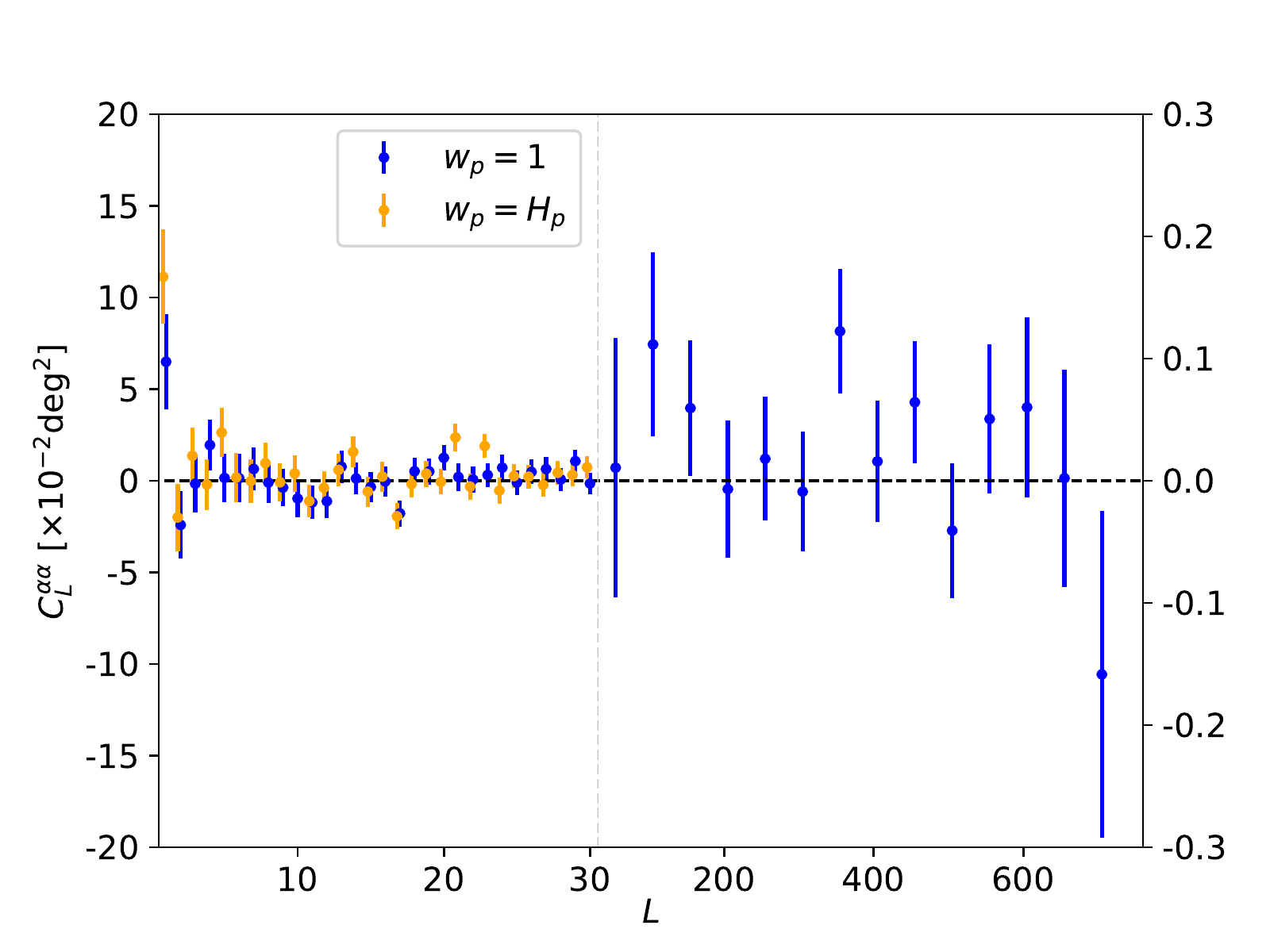}}
\mbox{\includegraphics[width=0.46\hsize]{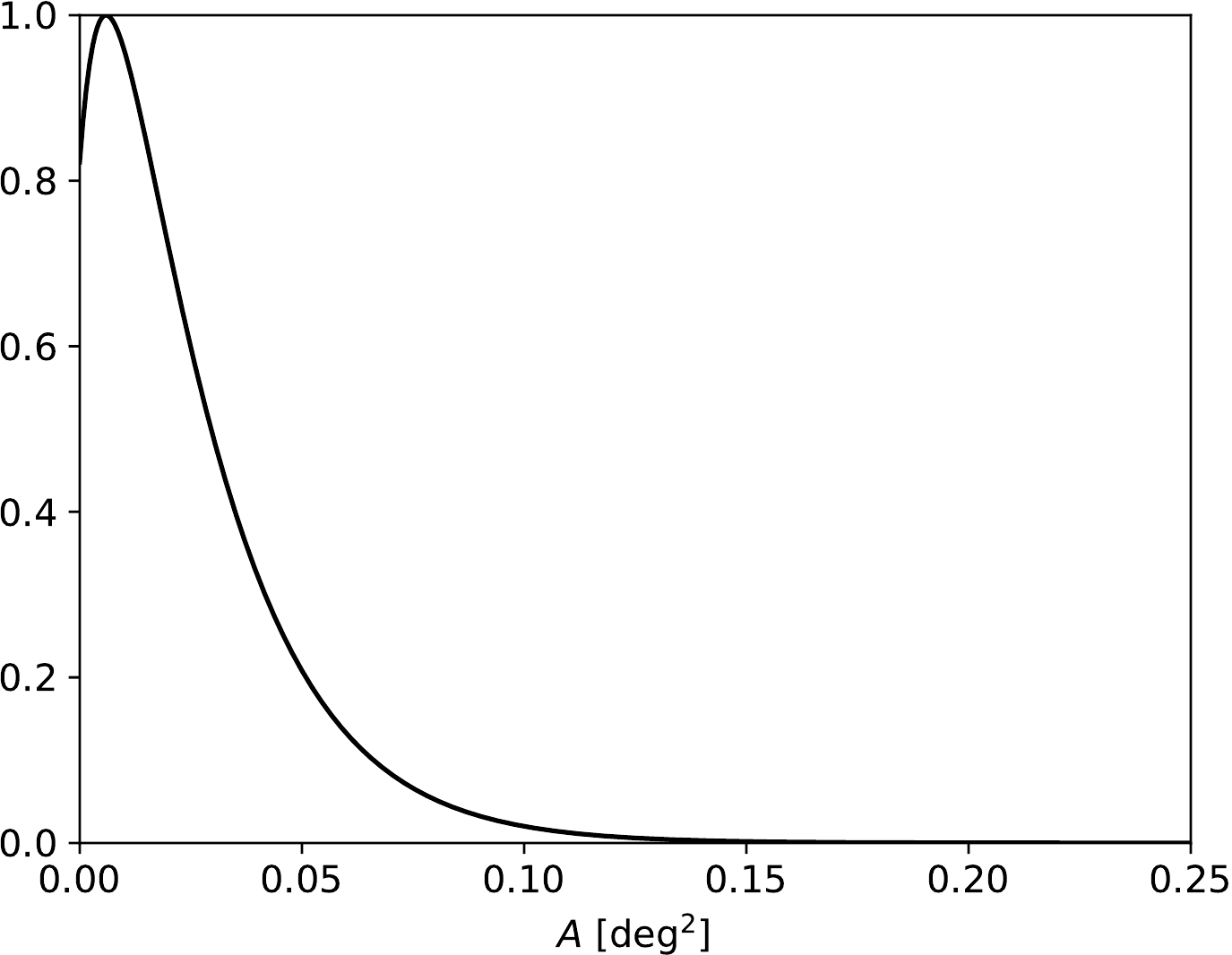}}
\end{center}
\caption{{\it Left}:
Power spectrum for $\alpha$. The vertical dashed grey line denotes the
boundary between our low $L$ and high $L$ reconstructions; note the differing
$y$-scale for low-$L$ compared to high-$L$. Uncertainties shown are standard
deviations of our set of null simulations; at low-$L$ the $C^{\alpha\alpha}_L$
are not Gaussian or symmetric, which is accounted for in our likelihood (see
Section~\ref{sec:lowl}). The power spectrum here justifies our use of the
small-angle approximation. {\it Right}: Posteriors for the amplitude ($A$) of a
scale-invariant power spectrum defined by Eq.~\eqref{eq:scaleinvariant}. The
constraint is mainly driven by the lowest $L$s, which is the reason for the
non-symmetric shape of the posterior.}
\label{fig:powerspectrum}
\end{figure}

The power spectrum at low $L$ and at high $L$ (binned) are shown together in
the left panel of Fig.~\ref{fig:powerspectrum}. Recall that at low $L$ the
spectrum is the mean of the auto-spectrum of our simulations subtracted from
the auto-spectrum of our low-$L$ $\alpha$-map. We find that this estimate is
consistent with the cross-correlation of $\alpha$ from half-mission~1 and
half-mission~2 data. The blue points in this figure use the uniform weighting
scheme, while the orange points use the hits-map weighting. At high $L$ the spectrum is
derived using cross-correlations only. We find good agreement with the
expectation of a null power spectrum over all $L$ (with the possible exception
of the $L=1$ mode, the discussion of which is left for
Section~\ref{sec:dipole}), and the smallness of the power spectrum justifies
our use of the small-angle approximation in Eq.~\eqref{eq:ebspectrum}.

\subsection{Constraints on a scale-invariant power spectrum}
\label{sec:scaleinvariant}

The posterior for the amplitude of a scale-invariant power spectrum is shown in
Fig.~\ref{fig:powerspectrum} (right). The high-$L$ likelihood prefers a
positive $A$ at less than the $2\sigma$ level, while the low-$L$ data are
consistent with $A=0$. However, they are both quite consistent with each other,
as can be seen by comparing the 95\,\% CL values for the low-$L$ likelihoods
and the full likelihood in Table~\ref{tab:95cl}. The full-$L$ constraint comes
from the combination of the low-$L$ and high-$L$ likelihoods, assuming no
correlation between the two. Due to the $L$ dependence of the model spectrum,
the likelihood is dominated by the lowest $L$s (a bluer spectrum would be more
constrained by high $L$ than the scale-invariant one). Note that the posterior is not
very Gaussian, since the $C_L$s follow a $\chi^2_{2L+1}$ distribution. We find
the constraint $A < 2.2\times10^{-5} \,(0.07\,[\mathrm{deg}^2])$ at 95\,\% CL.
This constraint is at a similar level to the systematic uncertainty of the
$\alpha$-monopole measurements, namely $0\fdg3$.  This suggests that, in the
absence of an improved absolute calibration scheme (see Ref.~\cite{Nati2017},
for an example of efforts in this direction), constraints on cosmic
birefringence from the next generation of CMB measurements will likely be
focused on searches for anisotropic $\alpha$.

While the noise level of \emph{Planck} polarization data is large compared to
the most recent ground-based experiments, it does have the distinct advantage
of measuring the largest scales. Our new constraints improve upon
the most stringent constraints available (see Table~\ref{tab:95cl})
and are an order magnitude smaller than previous results \cite{Gluscevic2012,
Ade2015}. Somewhat tighter constraints could be found using a joint low-$L$ and
high-$L$ analysis of data from \textit{Planck} combined with the BICEP2/Keck
Array.

\begin{table}[ht!]
\centering
\begin{tabular}{cccc}
\hline
\hline
Low $L$ $(w_p = H_p)$     & Low $L$ $(w_p = 1)$       & All $L$                   & BICEP2/Keck Array\\
\hline
$A \leq 2.4\times10^{-5}$ & $A \leq 1.9\times10^{-5}$ & $A \leq 2.2\times10^{-5}$ & $A \leq 3.3\times10^{-5}$\\
\hline
\end{tabular}
\caption{95\,\% CL upper limits on the amplitude $A$ of a scale-invariant power
spectrum. Here ``All $L$'' refers to the combination of our uniform weighting
low-$L$ and high-$L$ likelihoods. The last column comes from Ref.~\cite{B2K},
using polarization data from the BICEP2/Keck Array.}
\label{tab:95cl}
\end{table}

The difference between our low-$L$ likelihoods using two different weighting
schemes is attributable to residual foregrounds (explored in
Section~\ref{sec:foregrounds}) that are not accounted for in our simulations.
The hits-map weighting is more sensitive to these foregrounds compared to the
uniform weighting. In Section~\ref{sec:foregrounds} we derive a systematic
error for our amplitude, which primarily comes from residual polarized dust.

\subsection{The dipole}
\label{sec:dipole}

From the power spectrum, Fig.~\ref{fig:powerspectrum} (left panel), we see that
the dipole deviates the most from the expectation of a null power spectrum. We
quantify this by comparing $C_1$ for the data to the values in our simulations.
We find that only about 1.4\,\% of the simulations have a larger dipole than
the data.\footnote{For these results we include $\alpha$ obtained by our
$T$--$B$ estimator as well, although this has only a marginal influence on our results.}
This is the case for both the auto-correlation and cross-correlation of the
full-mission or half-mission data sets. It is therefore worth investigating
this signal further.

\begin{table}[ht!]
\centering
\begin{tabular}{lcc}
\hline
\hline
Method             & Amplitude $(A_1 = \sqrt{3C_1/4\pi})\, [{\rm deg}]$ & Direction $(l, b)\,[{\rm deg}]$\\
\hline
Uniform weighting  & $0.32 \pm 0.10$                                    & $(295, 17) \pm (22, 17)$\\
Hits-map weighting & $0.40 \pm 0.10$                                    & $(280, \phantom{1}1) \pm (15, 12)$\\
\hline
\end{tabular}
\caption{Mean posterior values and 68\,\% uncertainty levels for the amplitude
and direction of the dipole in $\alpha$. The corresponding 68\,\% radial
positional uncertainty around the best-fit direction is about $25^\circ$, with
a corresponding $p$-value of 1.4\,\%. The difference between both methods is
attributable to residual foregrounds (which are more apparent for the hits-map
weighting, see Section~\ref{sec:foregrounds}), as well as a significant
systematic effect.}
\label{tab:dipole}
\end{table}

\begin{figure}[ht!]
\begin{center}
\mbox{\includegraphics[width=0.49\hsize]{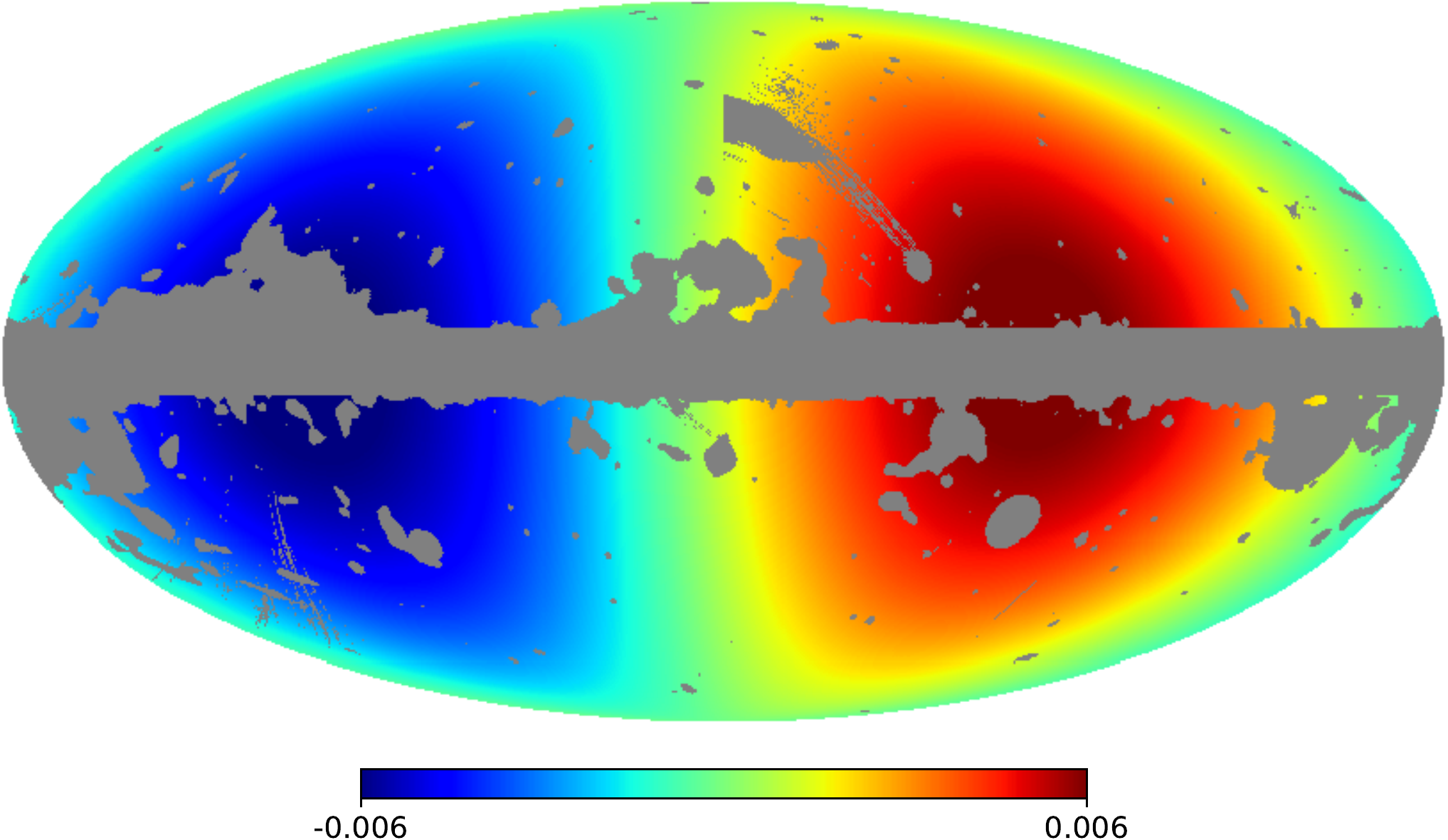}}
\mbox{\includegraphics[width=0.49\hsize]{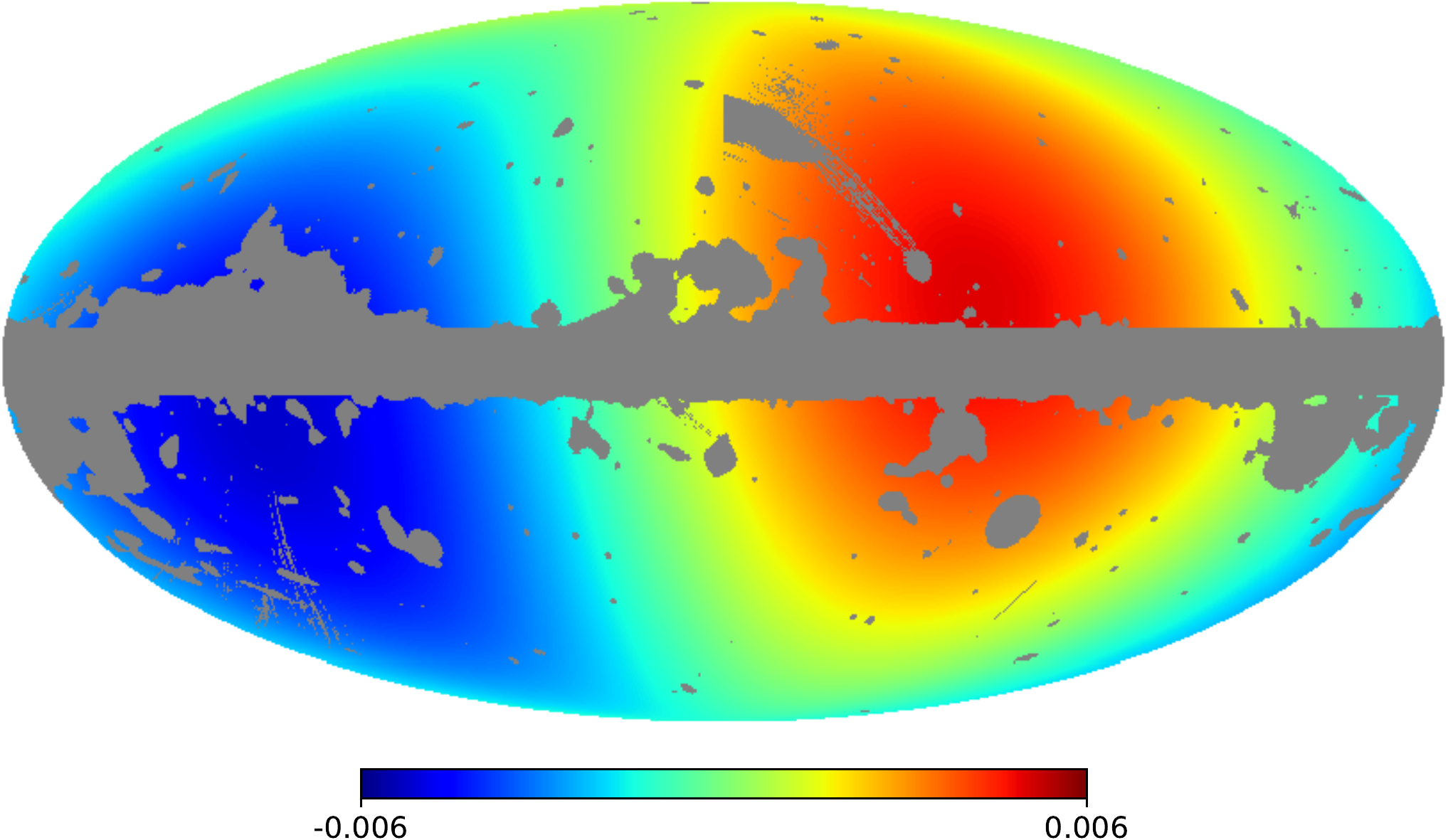}}
\end{center}
\caption{Best-fit dipole in $\alpha$ from the full-mission data using $w_p =
H_p$ (left), compared to the dipole from uniform weighting (right). The dipoles
are consistent, although the amplitude clearly decreases when using uniform
weighting.}
\label{fig:dipole}
\end{figure}

In the absence of a model, the values of $\alpha_{1M}$ are Gaussian distributed
with mean zero and variance given by $\left< C_1 \right>$, where the average is
taken over simulations with a null $\alpha$ dipole. We can convert this to an
amplitude and direction with a uniform prior on the $\alpha_{1M}$s,
parameterizing the dipole as
\begin{align}
  \alpha(\hat{\pmb{n}}) = A_1 \cos{\theta}.
  \label{eq:dipole}
\end{align}
Here $\theta$ is defined as the angle with respect to the best-fit direction
and $A_1 \equiv \sqrt{3C_1/4\pi}$. In Table~\ref{tab:dipole} we quote the mean
values of the posteriors and their corresponding 68\,\% uncertainties. We show
the best-fit dipoles in Fig.~\ref{fig:dipole} for our baseline results (right
panel) and using hits-map weighting (left panel). We explore in
Section~\ref{sec:foregrounds} the effect of residual foregrounds (not present
in our simulations) on the dipole and find that these significantly affect the
direction of the dipole.

If the dipole in $\alpha$ were to be physical (and not simply a statistical
fluctuation), then the overall signal could be fit by a sufficiently red
spectrum, which would have significant implications on the nature of the
sourcing pseudo-scalar (or vector) field. Alternatively, there could be a
genuinely preferred direction for the cosmic birefringence (i.e., a dipole that
is unconnected to a power spectrum). Either way, new
(preferably) all-sky polarization data with reduced noise levels compared to
\emph{Planck} are required to determine whether or not the dipole signal is
cosmological.

\subsection{\boldmath The $M=0$ quadrupole}
\label{sec:quadrupole}

While there is no evidence for a significantly large quadrupole in the data
(see Fig.~\ref{fig:powerspectrum}, left), there could be a special direction in
the $\alpha$ map that would show up solely in the $M=0$ mode. It is therefore
worth considering this mode specifically.

An $M=0$ quadrupole in an arbitrary direction would be related to our
coordinate $\alpha_{2M}$ values by a rotation with a Wigner $D$-matrix by
\begin{align}
  \alpha_{2M} &= \alpha'_{20} D^2_{M0} (l, b, 0).
  \label{eq:rotation}
\end{align}
Here $\alpha'_{20}$ is an $M=0$ quadrupole in a coordinate system pointing in
the direction $(l, b)$. With our simulations we generate a covariance, ${\sf
Q}$ from our simulations, which defines a likelihood of the form
\begin{align}
  \mathcal{L} &\propto \exp{\left[-\frac{1}{2}
  \left(\hat{\alpha}_{2M} - \alpha_{2M}(\alpha'_{20}, l, b)\right) {\sf Q}^{-1}
  \left(\hat{\alpha}_{2M} - \alpha_{2M}(\alpha'_{20}, l,
  b)\right)^\dagger\right]}.
  \label{eq:quadlike}
\end{align}
We then sample the likelihood using a Markov chain Monte Carlo (MCMC) algorithm
to obtain posteriors for the parameters $\alpha'_{20}$, $l$, and $b$.
Marginalizing over the direction we find $\alpha'_{20} = 0\fdg02 \pm 0\fdg21$,
which is very consistent with no special direction in the quadrupole. For this
same reason the direction is unconstrained.

\section{Systematic effects}
\label{sec:systematics}

In Section~\ref{sec:data} we mentioned several kinds of systematic effect that
are present in the \emph{Planck} polarization data that we have taken efforts
to avoid being sensitive to. These are: an uncertainty in the global orientation of the
PSBs; large-scale artefacts in the data; and an un-modelled correlated noise
component in the data. The first effect contributes a bias to a uniform
$\alpha$ and would thus cancel out in our search for anisotropic $\alpha$. The
second is mitigated by the use of high-pass-filtered data. The last effect is
minimized by using cross-correlations (between half-mission~1 and
half-mission~2 data) wherever possible. Our low-$L$ likelihood uses
auto-correlations, although we compared our power spectra to the corresponding
cross-correlation and found good consistency. However, this does not
definitively show that correlated noise is not a significant bias for our
results, and so we now further investigate other potential sources of
systematic effects.

\subsection{Foregrounds}
\label{sec:foregrounds}

Our estimator looks for parity violations in the CMB data in the hopes of
constraining a cosmological signal. Foregrounds contaminate this by
being large additive signals that do not source polarization in a way that is
invariant under parity transformations about our location, and hence can
corrupt the $\alpha$ signature.

We first test for this contamination by enlarging our baseline mask to cover
more of the Galactic plane. We take the \commonmask\ mask \cite{PlanckIX},
smoothed with a $200^\prime$ beam, then set all points below 0.9 to zero and all
others to 1, and finally multiply by the \commonmask\ mask again so as not to miss
any small masked areas. This decreases the sky fraction available from
\commonmask\ from $f_{\rm sky}=0.77$ to 0.69. This roughly 10\,\% decrease in
sky coverage leads to a roughly 40\,\% increase in the 95\,\% CL in the
amplitude of a scale-invariant power spectrum, due to increased sample
variance. Since the likelihood is dominated by low $L$ and is skewed to higher
values (see Fig.~\ref{fig:powerspectrum}, right panel), this result is still
quite consistent with our baseline result. However, we cannot entirely rule out
that foregrounds might be a significant systematic effect here. With that in mind,
it is nevertheless still the case that the increase in the 95\,\% CL limit is
consistent with \emph{no} detection of anisotropic $\alpha$.

As an additional test we can define an a posteriori mask to be zero everywhere
that the absolute value of the low-$L$ map in Fig.~\ref{fig:lowlmaps} (top
left) is greater than 0.15, masking the visually striking features. We find
that our power spectrum results remain consistent with the expectation of the
corresponding increased sample variance. In particular the dipole is consistent
in amplitude and direction with our baseline results from
Table~\ref{tab:dipole}.

Some foreground contaminants, such as dust, can produce $B$ modes that might
induce a cross-correlation signal between lensing and $\alpha$. We test for
this using the \emph{Planck} 2015 lensing maps \cite{PlanckXV} and perform the
cross-correlation with our low-$L$ data and simulations. Note that the
\emph{Planck} lensing maps contain no information for $L < 8$ \cite{PlanckXV},
so this test tells us nothing about the nature of the dipole in $\alpha$.
Nevertheless, we obtain a probability to exceed (PTE) the $\chi^2$ obtained
with the data (derived from simulations), of 25\,\%, consistent with no detection
of a cross-correlation.

\begin{table}[ht!]
\centering
\begin{tabular}{lcc}
\hline
\hline
Foreground  & Low-$L$ PTE [\%] & Dipole PTE [\%] \\
\hline
Dust        & \phantom{2}0.4   & 14\\
Synchrotron & 20\phantom{.0}   & 97\\
Free-free   & 13\phantom{.0}   & 96\\
\hline
\end{tabular}
\caption{Probability to exceed the $\chi^2$ obtained from the cross-correlation
of our $\alpha$ maps (or just the dipole) with the corresponding $\alpha^f$ map
(or just dipole) from each foreground of the data. We find a marginally
significant correlation with polarized dust, due to residual dust in the data
that is not accounted for in our simulations.}
\label{tab:foregrounds}
\end{table}

We also test for the presence of correlated polarized synchrotron, free-free,
and dust emission directly, by obtaining the \textit{Planck} $Q$, and $U$
foreground maps \cite{PlanckX} and propagating them through our analysis to
obtain maps of $\alpha$ (denoted $\alpha^f_{L M}$). That is, we use
Eq.~\eqref{eq:dirdependent} to fit for $\alpha$ in the foreground maps at the
location of peaks in the CMB $E$-mode map. These maps contain an unnormalized
estimate of the contribution from each foreground to $\alpha$ that is
correlated with the CMB. We then determine a PTE for the $\chi^2$ obtained with
the cross-correlation of $\alpha$ with $\alpha^f$, shown in
Table~\ref{tab:foregrounds}. We find marginally significant correlations with
dust only. In order to estimate the bias incurred from these foregrounds we
model the contamination as
\begin{align}
  \alpha_{LM} &= \tilde{\alpha}_{LM} + \sum_f \frac{C^{f\alpha}_L}{C^{ff}_L}
  \alpha^f_{LM}.
  \label{eq:contamination}
\end{align}
Here $\alpha_{LM}$ are the measured data, while $\tilde{\alpha}_{LM}$ are the
expected data coming just from the CMB, $C^{f\alpha}_L$ is the
cross-correlation of $\alpha$ obtained from the CMB and foreground, and
$C^{ff}_L$ is the auto-correlation of $\alpha$ obtained from the foreground
(with these quantities being estimated from the data). Propagating
$\tilde{\alpha}_{LM}$ to obtain 95\,\% CL values for the scale-invariant
amplitude leads to a \emph{decrease} of 0.7$\times10^{-5}$ (still consistent
with no detection). Note that this is the same amount of shift between our
low-$L$ likelihoods using hits-map weighting compared to uniform weighting; we
thus assign this value as a systematic error in our result, present because our
simulations do not contain residual foregrounds. When applied to the $L=1$ mode
specifically, we find that the dipole moves away from the Galactic plane by the
same amount as the difference between the hits-map weighted and uniform
weighted dipole (see Table~\ref{tab:dipole}), consistent with the presence of
residual correlated foregrounds; however, the amplitude remains largely
unchanged. We therefore assign a conservative systematic error of $0\fdg08$ to
the amplitude and $(5^\circ, 15^\circ)$ to the direction, substantially
impacting its significance.

\subsection{Point sources}
\label{sec:pointsources}

Although point sources in general add a source of bias to the 4-point function
\cite{Osborne2013nna}, they are not expected to contaminate the signal we are
looking for \cite{Gluscevic2012}, since they only contribute a parity-even
signal. Nevertheless we test the level of contamination by including a
point-source mask.

We consider the union of the {\it Planck\/} point source masks for polarization
from 100 to 353\,GHz. It turns out the vast majority of pixels masked by the
point source mask are \emph{already} masked by \commonmask. After degrading to
a common resolution of $N_{\rm side} = 1024$, there are 16 remaining pixels
(out of a potential 24{,}000) that are not also masked by \commonmask. It
should come as no surprise then that they therefore have a negligible effect on
our results and we thus consider point sources to be an unimportant systematic
for our analysis.

\subsection{Relative uncertainty on the PSB orientations}
\label{sec:relativeangle}

Although we are insensitive to a global rotation of the HFI detectors, we could
still in principle be sensitive to a relative angular separation between
individual PSBs. This is because, in the component separation process, different
frequencies (and thus PSBs) are used anisotropically, and thus a relative
difference in orientation of PSBs at different frequencies would appear as
anisotropic birefringence.

The relative upper limit on the PSB orientations is $0\fdg9$
\citep{Rosset2010}, however the dispersion of $\alpha$, as measured using
$E$--$B$ correlations between the HFI frequencies, is around $0\fdg2$
\cite{PlanckintXLVI}. We expect that such an anisotropy would appear
as a large-scale feature
characterizing the use of different frequencies
in the component-separation process. Thus we do not expect that this would
affect the search for a scale-invariant spectrum. Crudely speaking the use of
different frequencies varies mostly with latitude (though this depends upon the
method \cite{PlanckIX}), which is a pattern we do not see. In particular the
dipole is seen mainly in the Galactic plane, as opposed to the Galactic poles.
Therefore it seems unlikely that the excess in the dipole we see is due to
the relative uncertainty in PSB orientations compared to foregrounds; however,
a full characterization of such effects will be important for future studies.

\section{Conclusions}
\label{sec:conclusions}

We have estimated the anisotropy in the cosmological birefringence angle,
$\alpha$, with a novel map-space based method, using \emph{Planck} 2015
polarization data. Our results are consistent with no evidence for
parity-violating physics. We provide the most stringent constraints on the
anisotropy at \emph{large} angular scales and have constrained a
scale-invariant amplitude to be $A < [2.2\, (\mathrm{stat.})\, \pm 0.7\,
(\mathrm{syst.})]\times10^{-5}$ at 95\,\% CL.\footnote{Conservatively, one
should
take the full 95\,\% limit to be $2.9\times10^{-5} = 0.09 \,\mathrm{deg}^2$.} Here the systematic error comes
from estimating residual foregrounds (primarily dust) in the data. This implies
a constraint on dipolar and quadrupolar amplitudes to be $\sqrt{C_1/4\pi}
\lesssim 0\fdg2$ and $\sqrt{C_2/4\pi} \lesssim 0\fdg1$, respectively. These
constraints are, along with the newest results from the BICEP2/Keck Array, the
tightest limits on a scale-invariant power spectrum (see Table~\ref{tab:95cl}
for a direct comparison). We also search for special directions in $\alpha$,
finding that an $M=0$ quadrupole is constrained to be $\alpha_{20} = 0\fdg02
\pm 0\fdg21$, consistent with the null hypothesis. Our results are consistent
across four different component-separation methods and do not appear to be
significantly contaminated by point sources. We also find no significant
cross-correlation signal between our $\alpha$ maps and the \emph{Planck} 2015
lensing map.

One possible exception to the above conclusion is the dipole in $\alpha$ (whose
best-fit amplitude and direction can be found in Table~\ref{tab:dipole},
corresponding to a radial 68\,\% uncertainty on the direction of $25^\circ$),
which is somewhat large compared to null simulations, with an associated {\it
p}-value of 1.4\,\%. We find that the significance is insensitive to the use of
the auto-correlation of full-mission data or the cross-correlation of the
half-mission data. We do find that foreground contamination, coming primarily
from dust, biases the dipole in a significant way, pulling the direction toward
the Galactic plane, accounting for part of the signal. If, on the other hand,
some of the dipole is genuinely due to cosmic birefringence then this would
have significant implications for the form of the field and the source of its
fluctuations, necessitating a red spectrum or a specifically
direction-dependent birefringence. The model-space is vast and the significance
is low, and clearly partially contaminated by residual foregrounds, so we do
not speculate on what the physical source could be here. More sensitive
polarization data at large angular scales are required to settle the issue.

In Ref.~\cite{PlanckXLIX} it was determined that searches for a uniform angle
$\alpha$ are now dominated by systematic effects at the $0\fdg3$ level. Here we
find that constraints on the direction dependence of $\alpha$ are also at about
the $0\fdg3$ level, with no apparent dominant systematic effects limiting the
search in the near future. Therefore in the absence of an improved calibration
scheme for determining the orientation of the PSBs, future searches for
parity-violating physics of the form discussed here will likely be driven by
the pursuit of for anisotropic cosmic birefringence.

\acknowledgments

This work was supported by the Natural Sciences and Engineering Research
Council of Canada (NSERC).

\bibliographystyle{JHEP}

\bibliography{apj-jour,ani_bir}

\end{document}